\documentstyle[12pt,citesort]{article}

\def\mun{{\underline{m}}}
\def\nun{{\underline{n}}}
\def\kun{{\underline{k}}}

\catcode`\@=11

\newcount\hour
\newcount\minute
\newtoks\amorpm \hour=\time\divide\hour by 60\minute
=\time{\multiply\hour by 60 \global\advance\minute by-\hour}
\edef\standardtime{{\ifnum\hour<12 \global\amorpm={am}%
        \else\global\amorpm={pm}\advance\hour by-12 \fi
        \ifnum\hour=0 \hour=12 \fi
        \number\hour:\ifnum\minute<10
        0\fi\number\minute\the\amorpm}}
\edef\militarytime{\number\hour:\ifnum\minute<10
0\fi\number\minute}

\def\draftlabel#1{{\@bsphack\if@filesw {\let\thepage\relax
   \xdef\@gtempa{\write\@auxout{\string
      \newlabel{#1}{{\@currentlabel}{\thepage}}}}}\@gtempa
   \if@nobreak \ifvmode\nobreak\fi\fi\fi\@esphack}
        \gdef\@eqnlabel{#1}}
\def\@eqnlabel{}
\def\@vacuum{}
\def\marginnote#1{}
\def\draftmarginnote#1{\marginpar{\raggedright\scriptsize\tt#1}}
\def\draft{
        \pagestyle{plain}
        \overfullrule=2pt
        \oddsidemargin -.5truein
        \def\@oddhead{\sl \phantom{\today\quad\militarytime} \hfil
        \smash{\Large\sl DRAFT} \hfil \today\quad\militarytime}
        \let\@evenhead\@oddhead
        \let\label=\draftlabel
        \let\marginnote=\draftmarginnote
        \def\ps@empty{\let\@mkboth\@gobbletwo
        \def\@oddfoot{\hfil \smash{\Large\sl DRAFT} \hfil}
        \let\@evenfoot\@oddhead}
        \def\@eqnnum{(\theequation)\rlap{\kern\marginparsep\tt\@eqnlabel}%
        \global\let\@eqnlabel\@vacuum}  }

\newcommand{\rf}[1]{(\ref{#1})}
\renewcommand{\theequation}{\thesection.\arabic{equation}}
\renewcommand{\thefootnote}{\fnsymbol{footnote}}
\newcommand{\newsection}{    
\setcounter{equation}{0}\section}
\def\appendix#1{\addtocounter{section}{1}\setcounter{equation}{0}
\renewcommand{\thesection}{\Alph{section}}
\section*{Appendix \thesection\protect\indent \parbox[t]{10cm}{#1}}
\addcontentsline{toc}{section}{Appendix \thesection\ \ \ #1}}

\def\be{\begin{equation}}
\def\ee{\end{equation}}

\def\cI{{\cal I}}
\def\cJ{{\cal J}}

\def\prp{\pi^\oplus}
\def\prm{\pi^\ominus}

\def\PP{{\cal P}}

\def\x'{\mathaccent 19 x}
\def\th'{\mathaccent 19 \theta}
\def\barth'{\mathaccent 19 {\bar{\theta}}}

\def\k{k}
\def\l{l}

\def\ipr{{i^\prime}}
\def\jpr{{j^\prime}}
\def\kpr{{k^\prime}}
\def\lpr{{l^\prime}}

\def\del{\partial}

\def \vp {\varphi}

\def \L {\Lambda}
\def \a {\alpha}

\def \k {\kappa}

\def \s {\sigma}

\def \r {\rho}
\def \G {\Gamma}
\def \l {\lambda}
\def \m {\mu}
\def \g {\gamma}
\def \n {\nu}

\def \e#1 {{\rm e}^{#1}}

\def \vp {\varphi}

\def \H {{\hat H}}

\def \ha { { 1\over 2 }}

\def \ov {\over}

\def\Z'{\mathaccent 19 Z}

\def\mun{{\underline{m}}}
\def\nun{{\underline{n}}}
\def\kun{{\underline{k}}}

\def \adss {$AdS_5 \times S^5$\ }

\def \lc {light-cone\ }

\def \s { \sigma }

\def \vp {\varphi}

\def \vt {\theta}
 \def \a { \alpha}
\def \r {\rho}

\def \half {{1 \ov 2}}

\def \D {{\cal D}}

\def \DD {{\rm D}}

 \def \diag {{\rm diag}}

\def \adss {$AdS_5 \times S^5$\ }

\def \lc {light-cone\ }

\def\be{\begin{equation}}
\def\ee{\end{equation}}

\begin{document}


\begin{titlepage}
\begin{flushright}
hep-th/0202109\\
 FIAN/TD/02-04
\end{flushright}
\vspace{2 cm}

\begin{center}
{\LARGE
Exactly solvable model of
   superstring  \\[.3cm]
in plane wave  Ramond--Ramond  
background
 }\\[.2cm]
\vspace{1.1cm}
{\large R.R. Metsaev${}^{{\rm a}}$\footnote{\
E-mail: metsaev@lpi.ru}
and A.A. Tseytlin${}^{{\rm b,c}}$\footnote{
 E-mail: tseytlin.1@osu.edu} }

\vspace{18pt}

${}^{{\rm a\ }}${\it Department of Theoretical Physics, P.N.
Lebedev Physical Institute,\\ Leninsky prospect 53,  Moscow
119991, Russia }

\vspace{6pt}
  ${}^{{\rm b\ }}${\it
 Blackett Laboratory, Imperial College   \\
London, SW7 2BZ, U.K. \\
}

\vspace{6pt}
 ${}^{{\rm c\ }}${\it
 Department of Physics,
The Ohio State University  \\
Columbus, OH 43210-1106, USA\\
}

\end{center}

\vspace{2cm}

\begin{abstract}
We describe in detail the solution of type IIB superstring theory
in  the maximally supersymmetric plane-wave background with
constant null  Ramond-Ramond 5-form field strength. The
corresponding light-cone Green-Schwarz action found in
hep-th/0112044  is quadratic in both bosonic and fermionic
coordinates. We obtain  the light-cone  Hamiltonian
and  the string representation of the corresponding supersymmetry
algebra. The superstring Hamiltonian  has a ``harmonic-oscillator'' form
in  both the  string oscillator and the zero-mode parts
 and thus has  a discrete
spectrum. We analyze  the structure of the
 zero-mode sector  of the theory,   establishing the
precise correspondence  between the lowest-lying
``massless'' string states
and the  type IIB supergravity  fluctuation  modes
in the plane-wave  background.
The  zero-mode spectrum has certain similarity to  the
supergravity spectrum in $AdS_5 \times S^5$ background
of which the plane-wave background is a special limit.
We also compare  the  plane-wave string spectrum
with   expected form of the
  \lc gauge spectrum  of the $AdS_5 \times S^5$ superstring.

\end{abstract}

\end{titlepage}
\setcounter{page}{1}

\renewcommand{\thefootnote}{\arabic{footnote}}
\setcounter{footnote}{0}

\def \ci {\cite}
\def \g {\gamma}
\def \G {\Gamma}
\def \k {\kappa}
\def \l {\lambda}
\def \pw {plane-wave\ }
\def \foot {\footnote}

\def \u {x^{+}}
 \def \vv {x^{-}}
\def \S {{\tilde S}}
\def \om{\omega}
\def \hv{ {\hat v}} \def \hu{ {\hat u}} \def \hi {{\hat i} }
\def \A {\bar A}
\def \F {{\cal F}}
\def \bi{\bibitem}
\def \la {\label}

\def \lc { light-cone }
\def \pw { plane-wave  }

\def \vac {|0\rangle}
\def \T {{\rm T}}

\def\cI{{\cal I}}\def\cK{{\cal K}}

\def\cJ{{\cal J}}\def\cD{{\cal D}}

\def\PP{{\cal P}}

\def\R{{\scriptscriptstyle R}}
\def\L{{\scriptscriptstyle L}}

\def \mm {{\rm m}}

\def \vt {\theta}
\def \f {{\rm f}} \def \E {{\cal E}}
\def \H {{\rm h}}
\def \rhoh{\psi}
\def \Q {{\cal C}}

\def \D {{\cal D}}

\def \DD {{\rm D}}

\textwidth=16cm 
\textheight=22cm 

\newsection{Introduction}
The simplest gravitational plane wave  backgrounds
$$ ds^2 = 2 dx^+ dx^- + K(x^+,x^I) dx^+ dx^+ + dx^Idx^I \ , \ \ \ \ \
\ \ \ K= k_{IJ} x^I x^J$$
  supported by  a constant
NS-NS 3-form  background  provide examples of exactly solvable
(super)string models:  the  string action  becomes quadratic
in  the \lc gauge  $x^+ = p^+ \tau$ (see, e.g., \ci{all,rev,revc,revv}).
It was recently pointed out \ci{rrm} that this solvability
property
 is shared  also by
a  conformal   model describing type IIB  superstring
propagating  in a particular  \pw  metric supported by
a  {\it Ramond-Ramond}   5-form background \ci{bla}:
\be\la{bem}{ ds^2 = 2 dx^+dx^- - \f^2 x_I^2 dx^+ dx^+  + dx^I dx^I \
,   \ \ \ \ \ \ \  I=1,..., 8 \ ,  }  \ee
\be\label{bemm}
F_{+1234}= F_{+5678}= 2\f \  .     \ee
This  background  has several special properties.
It preserves the   maximal number of 32  supersymmetries \ci{bla},
and it is related by a special limit
  (boost along a  circle of $S^5$
combined with a rescaling of the coordinates and of
the radius or $\a'$) to the $AdS_5 \times S^5$  background \ci{blah}.
  The  exactly solvable
string theory corresponding  to \rf{bem}
may thus have some common features  with
a  much   more complicated
string theory on $AdS_5 \times S^5$
 whose \lc  action contains
non-trivial interaction terms \ci{mtfer,mtt}.

In the  present paper  which is an extension of \ci{rrm}  we  will
present  in  detail the solution of this R-R \pw   string
model. In particular,  we will explicitly identify  the massless
modes in  its spectrum with  small fluctuations of the type IIB
supergravity  fields in the   background \rf{bem}. The results
will   have  an obvious similarity  to those of \ci{kim} in the
case of \adss. In particular, a remarkable common feature of the
R-R  plane wave supermultiplets  and  the AdS supermultiplets  is
that   the massless  fields with different spins belonging to the
same supermultiplet  have, in general, different lowest energy
values.
 The same is true also  for
massive  supermultiplets.\foot{This is different from  what one finds
 in the case of the
 non-supersymmetric bosonic plane  wave backgrounds, where
 massless fields of  different spins have, as in the case
of the  flat space, the same lowest energy values.
This difference
is related to supersymmetry and not to the definition of
masslessness:  in both cases we use the same
definition  of massless fields
  based on so called $sim$ invariance (invariance under transformations
  of the original plane-wave algebra supplemented by the dilatation)
of the corresponding field equations \ci{bha,metp}.}

Let us first  recall the form of the \lc  gauge
Green-Schwarz   action
for the type IIB  superstring  in the background \rf{bem}.
This  action  was found in \ci{rrm} by using the supercoset method
of \ci{mt}, but there is  a  simple
short-cut argument relating the
presence of the fermionic  ``mass'' term   to the  form  of the
generalized spinor covariant derivative in type IIB supergravity.
In view of the special null Killing vector properties of the
background \rf{bem},\rf{bemm} it is possible to   argue that
the only non-vanishing  fermionic
contribution to the  type IIB superstring  action
in the standard \lc gauge
\be \la{gau}  x^+ = p^+ \tau\ , \ \ \ \
\ \ \ \  \G^{ +} \theta^{\cI}=0   \
\ee
comes from the  direct  covariantization
\be\label{bass}{{\cal L}_{2F}= i (\eta^{ab} \delta_{\cI\cJ} -
 \epsilon^{ab}
\rho_{3\cI\cJ}) \del_a
  x^{\mun } \bar \theta^\cI \Gamma_{ \mun} \cD_b
\theta^\cJ  \   }\ee
of the  quadratic  fermionic term  in the flat-space GS  \ci{GS}
action.
Here  $\theta^\cI$ ($\cI$=1,2)
are the  two real positive chirality  10-d  MW spinors
and $\rho_3=$diag(1,-1) (see Appendix for notation).
$  \cD$  is the generalized covariant
 derivative that
appears in the  Killing spinor equation
 (or gravitino transformation law)
in type IIB supergravity \ci{schwarz}:
acting on the  real  spinors
$\theta^\cI$ it has the form
(here we ignore the dilaton  and R-R scalar dependence
and rescale the R-R strengths by -2 compared to \ci{schwarz})
\be \la{derr}
\cD_a =
\del_a
 + { 1 \ov 4}
\del_a x^\mun \big[\     (\om_{ \m \n \mun}
 - \ha  H_{ \m  \n \mun} \rho_3) \G^{ \m  \n}
+
( { 1 \ov  3!} F_{\m\n\l} \G^{\m\n\l} \rho_1  +
 { 1 \ov 2\cdot  5!} F_{\m\n\l\r\k} \G^{ \m\n\l\r\k}\rho_0 ) \G_\mun
\  \big]
\ee
where  the $\r_s$-matrices in the $\cI,\cJ$ space
are the Pauli matrices
$ \rho_1 = \sigma_1$, $\rho_0 = i \sigma_2$.
In the \lc gauge \rf{gau} the
 non-zero contribution to \rf{bass} comes only from the term where
both the  ``external''  and ``internal''
$\del_a x^\mun  $ factors  in \rf{bass}
become  $p^+ \delta^\mun_+ \delta_a^0$.
As is well-known, in  the flat-space \lc  GS action
$\theta^1$ and $\theta^2$
become the right  and the left   moving  2-d fermions.
In the presence of the $F_5$-background \rf{bemm}
 the surviving  quadratic fermionic  term is
proportional to
$\theta^1  \G^{-} \G^{\m_1...\m_4} \theta^2  F_{+ \m_1...\m_4 }$.
While  in the case of an   NS-NS 3-form background
the  fermionic interaction term has  a chiral 2-d form
($\rho_3$ is diagonal),
in the case of a
R-R
background
one gets   a   non-chiral  2-d ``mass-term'' structure
($\rho_1$  and $\rho_0$ are off-diagonal)
out of the interaction term in $\cD_a$ in \rf{bass},\rf{derr}.

The resulting quadratic   \lc  action  \ci{rrm}
 can be
written, like the  flat-space  GS action,  in
a 2-d spinor form  and
describes 8 free massive
2-d  scalars and $8$  free
massive Majorana  2-d  fermionic fields $\psi = (\theta^1,\theta^2)$
propagating in  flat 2-d  world-sheet
\be\la{laa}
 {\cal L}= {\cal L}_B  + {\cal L}_F  \ , \ \ \ \ \ \ \ \ \ \
{\cal L}_B = \ha(  \del_+ x^I \del_- x^I  -  \mm^2  x^2_I)   \ ,
\ \ \ \ \ \  \    \mm \equiv p^+ \f \ ,
\ee
  \be \la{act} {\cal L}_F =
{\rm i}  ( \theta^1\bar{\gamma}^- \partial_+  \theta^1   +
\theta^2 \bar{\gamma}^-\partial_-  \theta^2 -2 \mm    \theta^1
\bar{\gamma}^- \Pi \theta^2 )  \, ,  \ \ \ \ \ \    \bar \gamma^+
\theta^{\cI}=0  \ .  \ee
Here   $\del_\pm= \del_0 \pm \del_1$ and
we absorbed   one factor of $p^+$ into $\theta^\cI$.
We use the  spinor notation of \ci{rrm}, i.e.
$ \g^m , \
\bar \g^m  $ are  the $16 \times 16$ Dirac matrices
which are the off-diagonal parts of $32 \times 32 $   matrices  $\G^m$.
The matrix  $\Pi$ in the mass term  ($\Pi^2 =1$)
is the product of  four $\gamma$-matrices (see Appendix)
which   originates from
$\G^{\m_1...\m_4}   F_{+ \m_1...\m_4 }$ in
 \rf{bass},\rf{derr}.

In section 2.1 we shall review the solution of the classical
equations corresponding to the \lc  gauge action \rf{laa},\rf{act}
and then  (in section 2.2) perform
  the  straightforward  canonical quantization of this
 quadratic system  already sketched  in \ci{rrm}.
In section 2.3  we shall present the \lc string realization
of the basic symmetry superalgebra of the  \pw background.
 We shall  then use this  superalgebra to fix
the vacuum-energy (``normal-ordering'')
 constant in the zero-mode sector (section 2.4).
As we shall explain, the choice of the fermionic zero-mode vacuum
is not unique with different
(physically equivalent) choices
depending on how one  decides to describe
 the  representation
of the corresponding Clifford algebra. In particular, we
 note that a
 choice that leads to zero vacuum energy constant
breaks the $SO(8)$  global symmetry down to
$SO(4) \times SO'(4)$ (which is in fact the true symmetry
of the \pw background \rf{bem},\rf{bemm})  but
is not the one that has a smooth flat-space limit.

In section 3  we shall  determine the spectrum of fluctuations
of  type IIB  supergravity   expanded near the \pw
background \rf{bem},\rf{bemm}.
Section 3.1 will contain some general remarks on
 solutions of  massless Klein-Gordon-type
 equations in the \pw metric \rf{bemm}.
The bosonic (scalar, 2-form, graviton and 4-form field)
spectra will be  found in section 3.2. The spin 1/2 and spin 3/2
cases will be  analyzed in section 3.3.
 Our analysis will be  similar to the
one carried out in \ci{kim} in the case of the \adss background.
As a result, we will be  able to  give a space-time
interpretation  to the ``massless'' (zero-mode)
sector of the string theory.  The discreteness of the supergravity
part of the  \lc energy spectrum will follow from the  condition of
square-integrability of the solutions of the corresponding
wave equations at fixed $p^+$.
In section 3.4  we will summarize the results for the bosonic and fermionic spectra in  the two tables and then explain
 how the corresponding physical modes can be interpreted
 as components  of
 a single  scalar   type IIB superfield satisfying a  massless
(dilatation-invariant) equation in \lc superspace.

In the concluding section 4  we shall  make some comments on the
parameters and possible  limits  of the \pw string theory, and also
compare it with the  expected  form
of the
 \lc string theory spectrum in \adss background.

Our index and spinor notation and definitions
as well as  some useful relations
will be  given   in Appendix.

\newsection{Canonical quantization }

\subsection{Solution of classical equations}
The equations of motion following from   \rf{laa},\rf{act} take the form:
\be \la{xxx}   \del_+ \del_- x^I  + \mm^2 x^I=0\,  ,  \ee
 \be
\label{freeq1}
\partial_+ \theta^1  -
\mm \Pi \theta^2 =0\,,\qquad \partial_- \theta^2  +
\mm \Pi \theta^1 =0\,.\ee
 The parameter
$\f$ in \rf{bem}  which has dimension of mass
 can be absorbed into rescaling of $x^+,x^-$, i.e. set to a
 given value.\foot{Note also that since  the
generator $P^+$ commutes with all other
 generators of the
plane wave superalgebra we could fix
  $p^+$ to take some specific non-vanishing value.
  In what follows
 we shall  $p^+$  arbitrary.}
We shall choose   the length of $\sigma$-interval to be 1.
The flat space limit  corresponds to $\mm \to 0$.

As follows from the structure of the covariant
string action corresponding to the background \rf{bem},\rf{bemm}
one can  absorb the dependence on the string tension
into  the following rescaling of the coordinates\foot{After the rescaling
$x^-,x^I$ will be dimensionless (like $\tau$ and $\sigma$)
but $x^+$ (and $\a' p^+$)  will have dimension of length.}
$x^- \to  2\pi \a'  x^-, \ x^I \to  (2\pi \a')^{1/2} x^I, \
\theta^\cI \to (2\pi \a')^{1/2} \theta^{\cI}$
with $x^+$ unchanged.
Then all  one needs to do  to restore
the dependence on the string tension
 is the following rescaling  of $p^+$
 \be\la{resca}
p^+ \ \ \ \to \ \  2\pi \a'  p^+  \ .  \ee
In particular,  $\mm \to \mm=2\pi \a'  p^+  \f$.

The general solutions to \rf{xxx},\rf{freeq1}
satisfying the  closed string boundary conditions
\be x^I(\sigma+1,\tau) = x^I(\sigma,\tau)\,,\qquad
\theta(\sigma+1,\tau)=\theta(\sigma,\tau)\,,\qquad \qquad 0\leq
\sigma \leq 1\ , \ee
are found to be
\be\label{xIsol} x^I(\sigma,\tau) = \cos \mm \tau\, x_0^I +\mm^{-1} \sin \mm
\tau\, p_0^I + {\rm i}\sum_{n=\!\!\!\!/\,\, 0
}\frac{1}{\omega_n}\Bigl( \varphi_n^1(\sigma,\tau) \alpha_n^{1I}
+ \varphi_n^2(\sigma,\tau) \alpha_n^{2I}\Bigr) \ee
\be \label{th1sol}\theta^1(\sigma,\tau) =\cos\mm\tau\, \theta_0^1
+\sin\mm\tau\  \Pi\theta_0^2+ \sum_{n=\!\!\!\!/\,\, 0}c_n\Bigl(
\varphi_n^1(\sigma,\tau) \theta^1_n + {\rm
i}{\textstyle{\omega_n-k_n\ov \mm}}\varphi_n^2(\sigma,\tau)\Pi \theta_n^2\Bigr) \ee
\be\label{th2sol} \theta^2(\sigma,\tau) =\cos\mm\tau\, \theta_0^2 -
\sin\mm\tau\ \Pi\theta_0^1+ \sum_{n=\!\!\!\!/\,\, 0} c_n\Bigl(
\varphi_n^2(\sigma,\tau) \theta^2_n - {\rm
i}{\textstyle{\omega_n-k_n\ov \mm}}\varphi_n^1(\sigma,\tau)\Pi \theta_n^1\Bigr) \ee
where the basis functions $\varphi^{1,2}_n(\sigma,\tau)$ are
\be\varphi^1_n(\sigma,\tau) = \exp(-{\rm i}(\omega_n \tau
-k_n\sigma))\,,\qquad \varphi^2_n(\sigma,\tau) = \exp(-{\rm
i}(\omega_n \tau + k_n\sigma))\ee
and

\be \omega_n =\sqrt{k_n^2 + \mm^2 },\quad n>0\ ;\qquad
 \omega_n
=-\sqrt{k_n^2 + \mm^2 },\quad  n < 0\ ;\ee \be \label{cn} k_n
\equiv 2\pi n \ , \ \ \ \ \ \ \ \ \ \ \ \  \  c_n
=\frac{1}{\sqrt{1+({\omega_n -k_n\ov \mm})^2}}\,, \qquad n=\pm
1,\pm 2,\ldots \ .  \ee
 The canonical momentum
$\PP^I =
\dot{x}^I$ takes the form

\be \PP^I(\sigma,\tau)= \cos \mm\tau\, p_0^I - \mm\sin\mm \tau\,
x_0^I + \sum_{n=\!\!\!\!/\,\, 0 }\Bigl( \varphi_n^1(\sigma,\tau)
\alpha_n^{1I} + \varphi_n^2(\sigma,\tau) \alpha_n^{2I}\Bigr) \ee
 The fermionic momenta given by $-{\rm i}\bar{\gamma}^-
\theta^\cI$  imply  that there are the second class constraints
which should be treated following the standard Dirac procedure
(see, e.g., \ci{rrm}).

The coordinate $x^-$ satisfies  the equation

\be    p^+\x'^- + \PP^I\x'^I + {\rm i} ( \theta^1\bar{\gamma}^-
\th'{}^1 +\theta^2 \bar{\gamma}^- \th'{}^2)=0
\ ,   \ee
{\
which leads to  the
constraint

\be \label{con3} \int d \sigma [ \PP^I\x'^I + {\rm i} (
\theta^1\bar{\gamma}^- \th'{}^1 +\theta^2 \bar{\gamma}^-
\th'{}^2)] =0\ . \ee
We get the following classical
Poisson-Dirac brackets

\be\label{aacom} [p_0^I,x_0^J]_{P.B.}=\delta^{IJ}\,,\qquad
[\alpha_m^{\cI I},\alpha_n^{\cJ J}]_{P.B.} =\frac{\rm i}{2}\omega_m
\delta_{m+n,0}\delta^{IJ}\delta^{\cI\cJ} \ ,  \ee

\be\label{ttcom1}\{\theta_m^{\cI\alpha},\theta_n^{\cJ\beta}\}_{P.B.}
=\frac{\rm i}{4 }(\gamma^+)^{\alpha\beta}
\delta^{\cI\cJ}\delta_{m + n,0}\ . \ee
 The matrix  $\gamma^+$ in
\rf{ttcom1} is reflecting the fact that we are using  the \lc
gauge constrained fermionic coordinates, $\bar \gamma^+
\theta^\cI=0$. The coefficients $c_n$ \rf{cn} are chosen so
that the Fourier modes of the
fermionic coordinates satisfy the
standard Poisson-Dirac brackets \rf{ttcom1}.

\subsection{Quantization and space of states}

 We
 can now quantize  2-d fields   $x^I$ and $\theta^\cI$
by promoting as usual  the   coordinates and momenta or
the Fourier components appearing in
\rf{xIsol},\rf{th1sol},\rf{th2sol} to operators
and replacing  the
 classical  Poisson (anti)brackets \rf{aacom}\rf{ttcom1}
by the   equal-time  (anti)commutators
of  quantum coordinates and momenta
according to the rules $\{.\,, .
\}_{P.B.}\rightarrow {\rm i} \{.\,, .\}_{quant}$, $[.\,, .
]_{P.B.}\rightarrow {\rm i} [.\,, .]_{quant}$.
This gives ($m,n = \pm 1, \pm 2, ...$)

 \be \label{comrel1} [p_0^I,x_0^J]=-{\rm i}\delta^{IJ}\,,\qquad
[\alpha_m^{\cI I},\alpha_n^{\cJ J}] =\frac{1}{2}\omega_m
\delta_{m+n,0}\delta^{IJ}\delta^{\cI\cJ}\,,\ee

\be\label{comrel3} \{\theta_0^{\cI\alpha},\theta_0^{\cJ\beta}\}
=\frac{1}{4 }(\gamma^+)^{\alpha\beta} \delta^{\cI\cJ}\ , \ \ \ \ \ \
 \{\theta_m^{\cI\alpha},\theta_n^{\cJ\beta}\}
=\frac{1}{4 }(\gamma^+)^{\alpha\beta} \delta^{\cI\cJ}\delta_{m +
n,0}\,.\ee

The \lc  superstring Hamiltonian  is

\be   H\equiv  -  P^-  \ , \ee \be\label{ham}  H = { 1 \ov p^+}
 \int^1_0  {d \sigma} \big[  \frac{1}{2 } (\PP_I^2 +
\x'_I^2+ \mm^2 x_I^2) + 2{\rm i}\mm \theta^1\bar{\gamma}^-\Pi
\theta^2 - {\rm i}(\theta^1 \bar{\gamma}^-  \th'^1
-\theta^2\bar{\gamma}^-\th'^2)\big] \ . \ee Using the fermionic
equations  of motion it can be rewritten in the form
\be\label{pmin3} H ={ 1 \ov p^+} \int {d\s}
 \ \big[ \frac{1}{2 } (\PP_I^2 + \x'_I^2+
\mm^2 x_I^2) +{\rm i}(\theta^1 \bar{\gamma}^- \dot{\theta}{}^1
+\theta^2\bar{\gamma}^-\dot{\theta}{}^2) \big] \ .
\ee
Plugging  in   the above expressions  for the
coordinates and momenta we  can represent  the resulting
\lc energy operator as
\be\la{ttt}
  H = E_0 + E^1 +E^2 \ , \ee
where $E_0$ is  the  contribution of the
zero modes  and
$E^{1},E^2$ are the  contributions of  the string oscillation modes

\be \la{eee}E_0 = \frac{1}{2p^+}(p_0^2 +\mm^2 x_0^2) + 2{\rm i}\f\
\theta_0^1\bar{\gamma}^-\Pi\theta_0^2 \ , \ee

\be \la{hhh} E^\cI = \frac{1}{p^+}\sum_{n=\!\!\!\!/\,\,
0}(\alpha_{-n}^{\cI I} \alpha_n^{\cI I}+
\omega_n\theta_{-n}^\cI\bar{\gamma}^-\theta_n^\cI ) \ , \ \ \ \ \
\ \ \ \cI=1,2  \  . \ee The constraint \rf{con3} takes the form

\be N^1=N^2  \ , \ \ \ \ \ \ \ \ \ \ \
 N^\cI \equiv \sum_{n=\!\!\!\!/\,\,0}
(\frac{k_n}{\omega_n}\alpha_{-n}^{\cI I}\alpha_n^{\cI I} +
k_n\theta_{-n}^\cI\bar{\gamma}^-\theta_n^\cI )\,. \ee Let us
introduce the  following basis of creation and annihilation
operators

\be \label{bososc} a_0^{I} = \frac{1}{\sqrt{2\mm}}(p_0^I + {\rm
i}\mm x_0^I)\,,\qquad \bar{a}_0^I =\frac{1}{\sqrt{2\mm}}( p_0^I -
{\rm i}\mm x_0^I)\,, \ee

\be \alpha_{-n}^{\cI I }
=\sqrt{\frac{\omega_n}{2}}\,a_n^{\cI I}\,,\qquad
 \alpha_n^{\cI I } =\sqrt{ \frac{\omega_n}{2}}\,\bar{a}_n^
 {\cI I} \,,\qquad n =1,2,\ldots \ee

\be \la{nss}
\theta_0 =   \frac{1}{\sqrt{2}} (\theta^1_0 + i \theta^2_0)
\ , \ \ \ \ \ \ \ \ \
\bar  \theta_0 =  \frac{1}{\sqrt{2}}(\theta^1_0 - i \theta^2_0)
 \ , \ee
\be \la{ddo}
\theta^\cI_{-n}\equiv \frac{1}{\sqrt{2}}\eta^\cI_n  \ ,
\qquad \theta^\cI_n\equiv
\frac{1}{\sqrt{2}}\bar{\eta}^\cI_n\,,\qquad n=1,2,\ldots  \ee in
terms of which  the commutation relations
\rf{comrel1},\rf{comrel3} take  the form
\be \la{huh}
[\bar{a}_0^I, a_0^J]=\delta^{IJ} \ , \ \ \ \ \ \ \
 [\bar{a}_m^{\cI I},a_n^{\cJ
J}]=\delta_{m,n}\delta^{IJ}\delta^{\cI\cJ}\,,\ee \be \la{pip}
\{\bar{\theta}_0^\alpha,\theta_0^\beta\}
=\frac{1}{4}(\gamma^+)^{\alpha\beta}\,,\qquad
 \{\bar{\eta}_m^{\cI}{}^\alpha,\eta_n^{\cJ}{}^\beta\}
=\frac{1}{2}(\gamma^+)^{\alpha\beta}\delta_{m,n}\delta^{\cI\cJ} \ . \ee
Here $\a=
1,...,16$,  and the spinors  are subject to the $\bar \gamma^+
\theta^\cI_0 =0, \bar \gamma^+ \eta^\cI_n =0$ constraint.

In this basis  the \lc energy   operator \rf{ttt}
  becomes the sum  of $E_0$,  $E^1$ and $E^2$  where
\be\label{ezer} E_0 = \f \E_0 \ , \ \ \ \ \ \ \ \   \ \ \ \ \
\E_0 =  a_0^I\bar{a}_0^I +
2\bar{\theta}_0^{}\bar{\gamma}^-\Pi\theta_0^{} + 4   \ , \ee \be
\la{eeee} E^\cI = {1 \ov p^+} \sum_{n=1}^\infty {\omega_n}
(a_n^{\cI I} \bar{a}_n^{\cI I}+
\eta_n^\cI\bar{\gamma}^-\bar{\eta}_n^\cI )\,.
  \ee We have normal-ordered the bosonic zero modes in $\E_0$
(getting extra  term $ \ha \times 8 =4$)
 and
both the bosonic and fermionic operators
 in $E^\cI$ (here the normal-ordering constants
cancel out
 as there are
equal numbers of bosonic and fermionic oscillators).
Note that because of
the relation Tr$(\gamma^+\bar \gamma^-\Pi)=0$
the contribution of the fermionic
zero  modes in  \rf{ezer}
does not depend on ordering of $\theta_0$   and $\bar \theta_0$.

To restore the dependence on $\a'$ we  need to rescale $p^+$ as in
\rf{resca}. The explicit form of the \lc Hamiltonian is then
\be\label{zer} H = \f ( a_0^I\bar{a}_0^I +
2\bar{\theta}_0^{}\bar{\gamma}^-\Pi\theta_0^{} + 4) +  {1 \ov \a'
p^+ } \sum_{\cI=1,2} \sum_{n=1}^\infty {\textstyle{ \sqrt{ {n^2}
+ (\a' p^+  \f)^2 }}} \    (a_n^{\cI I} \bar{a}_n^{\cI I}+
\eta_n^\cI\bar{\gamma}^-\bar{\eta}_n^\cI ) \ .
 \ee
Note that the energy thus depends on the two parameters of
 mass  dimension 1:  the curvature (or R-R field) scale
 $\f$  and the string scale  $ (p^+\a')^{-1}$.
The flat-space limit corresponds to $\f=0$
(the zero-mode part recovers  its flat-space form
${ p^2_I\ov 2 p^+}$ as in the case of the standard harmonic
oscillator, cf. section 3.1).

The vacuum  state  is the direct product of a  zero-mode
vacuum and the Fock vacuum for string oscillation modes, i.e. it
is defined  by
\be\label{vacdef} \bar{a}_0^I|0\rangle=0\,,
\qquad \bar{\theta}_0^\alpha|0\rangle=0\,,\qquad \bar{a}_n^{\cI
I}|0\rangle=0\,,\qquad \bar{\eta}_n^{\cI\alpha}|0\rangle =0
\ , \ \ \ \ n=1,2,... \ .  \ee
Generic  Fock space vectors are then
built up  in terms of  products of  creation
operators $a_0^I$, $a_n^{\cI I}$, $\theta_0^\alpha$,
$\eta_n^{\cI,\alpha}$ acting on the vacuum

\be |\Phi\rangle =\Phi(a_0\,, a_n\,,  \theta_0\,,
\eta_n)|0\rangle\,.\ee The subspace of physical states  is
obtained by imposing the constraint
\be\la{cons}
 N^1|\Phi_{phys}\rangle
=N^2|\Phi_{phys}\rangle \  , \ \ \ \ \ \ \ \ \
 N^\cI = \sum_{n=1}^\infty k_n(a_n^{\cI I}\bar{a}_n^{\cI
I} + \eta_n^\cI\bar{\gamma}^-\bar{\eta}_n^\cI) \ . \ee
Note that in contrast to the  flat space
case here  $E^{\cI}\not= N^{\cI}$.

Let us now make  few  remarks about the global symmetry of the
above
expressions.
While the metric \rf{bem}  and the bosonic part of the string action
\rf{laa}  have $SO(8)$ symmetry, the  5-form
background \rf{bemm} and  thus  the  fermionic
part of the classical action \rf{act}  is invariant
only  under $SO(4) \times SO'(4)$.
The
contribution of the string oscillators to the  Hamiltonian \rf{eeee} is
$SO(8)$ invariant,  but this  invariance is broken down to
$SO(4) \times SO'(4)$
 by  the contribution
 of the fermionic zero modes in \rf{ezer}.
In general, the amount of global symmetry
of the zero-mode Hamiltonian depends on  the
definition of the fermionic
creation and annihilation  operators, i.e.
on the definition of the  zero-mode vacuum.
With the definition  used in \rf{nss}
the vacuum  \rf{vacdef}  preserves
$SO(8)$ symmetry, but
the  fermionic part of the
zero-mode Hamiltonian \rf{ezer} is not  $SO(8)$ invariant.
 One can  instead  introduce  another  set of fermionic
creation/annihilation
operators,  i.e. use another definition of  the
fermionic zero-mode vacuum,
which preserves  only the $SO(4)\times SO'(4)$ invariance,
but which formally restores
the  $SO(8)$ invariance of the zero-mode Hamiltonian
(see section  2.4 below).
In any case,  the
$SO(8)$ invariance  is
broken  down to $SO(4)\times SO'(4)$
 not only in  the  fermionic  zero mode sector,
 but also explicitly
by  the  string-mode contributions
to  the  dynamical supercharges discussed  in section  2.3.

\subsection{Light cone string realization of the  supersymmetry algebra}

In general,  the choice of the light-cone
gauge  spoils part of   manifest global symmetries,  and  in order to
demonstrate that these  global invariances are still present, one
needs to  find  the (bosonic and fermionic) Noether charges
  that generate
them. These charges play   a crucial role  in formulating
superstring field theory in the light-cone gauge  in flat space
\ci{GSB,GS1} and are of equal  importance in the present \pw
context (see also \ci{rrm}).

In  the light-cone formalism, the generators (charges)  of the
basic superalgebra  can be split into the
 kinematical generators $
P^+,\,\, P^I,\,\, J^{+I},\,\, J^{ij},\,\, J^{\ipr\jpr},\,\,
Q^{+},\,\, \bar{Q}^+, $ and the  dynamical generators $
P^-,\,\, Q^-,\,\, \bar{Q}^-$ (here  $I=(i,i') $, $i=1,2,3,4;$ \ $i'=5,6,7,8$).
\foot{At point $x^+=p^+\tau=0$ the
kinematical generators in the superfield realization are
quadratic in the  physical string fields, while the dynamical
generators receive higher-order interaction-dependent
corrections.} It is important to  find  a  free (quadratic) field
representation for  the generators of the basic superalgebra.
The kinematical generators  which effectively depend only on the
zero modes  are\foot{We define $\theta \equiv {1 \ov \sqrt 2}
(\theta^1 + i \theta^2)$, $\bar{\theta} \equiv {1 \ov \sqrt 2}
(\theta^1 - i \theta^2)$.} \be\label{kin1} P^+  = p^+ \,,\qquad\
\ \ \ \
 P^I = \int d \s (  \cos \f x^+\  \PP^I  +\f \sin \f x^+\ x^I p^+) \,, \ee
\be\label{kin2}
 J^{+I} = \int d \s ( \f^{-1} {\sin \f x^+}\ \PP^I -\cos \f  x^+\ x^I p^+) \,,\ee
\be \label{kin3}  Q^+ = 2\sqrt{p^+}\int d \s
\bar{\gamma}^-e^{{\rm i}\f x^+ \Pi}\theta \,,\qquad
 \bar{Q}^+ = 2\sqrt{p^+}\int d\s \bar{\gamma}^-e^{-{\rm i}\f x^+
\Pi}\bar{\theta}\,. \ee
The remaining kinematical charges $J^{IJ}=( J^{ij},J^{\ipr\jpr })$
have non-zero components  which depend on all
 string modes are
\be\label{kinij}  J^{ij} = \int d \s ( x^i \PP^j - x^j\PP^i  -{\rm i}
\bar{\theta}\bar{\gamma}^-\gamma^{ij}\theta ) \,, \ \ \ \ \ \
 J^{\ipr\jpr } = \int d \s  ( x^\ipr \PP^\jpr - x^\jpr\PP^\ipr  -{\rm
i} \bar{\theta}\bar{\gamma}^-\gamma^{\ipr\jpr}\theta) \,. \ee The
dynamical charge $P^-$ is given by \rf{ham}, while the
supercharges $Q^-$ and $\bar{Q}^-$ are given by ($ Q,\bar Q = { 1
\ov \sqrt 2 } (Q^{1} \pm
 i Q^{2}) $)
\be \label{dyncha3} Q^{-1} = \frac{2}{\sqrt{p^+}}\int d \s [
(\PP^I - \x'{}^I)\bar{\gamma}^I \theta^1 - \mm x^I\bar{\gamma}^I
\Pi\theta^2 ]\,, \ee \be
 Q^{-2} = \frac{2}{\sqrt{p^+}}\int d\s [ (\PP^I +\x'{}^I)\bar{\gamma}^I
\theta^2 + \mm x^I\bar{\gamma}^I \Pi\theta^1] \,,\ee The
derivation of these supercharges was given in \ci{rrm}.

Using the mode expansions of section 2.1 in \rf{kin1},\rf{kin3}
we get
by\footnote{While transforming the generators $J^{\mu\nu}$ \rf{kin2},\rf{kinij}
to the form given in \rf{kin11},\rf{kinij2} we multiply them
by factor $+{\rm i}$.}
 \be\label{kin11} P^+  = p^+ \,,\qquad
 P^I =  p_0^I   \,,
\qquad  J^{+I} =  - {\rm i}  x_0^Ip^+  \,,\ee \be  Q^+ =
2\sqrt{p^+} \bar{\gamma}^-\theta_0 \,,\qquad
 \bar{Q}^+ = 2\sqrt{p^+} \bar{\gamma}^-\bar{\theta}_0\,. \ee
The charges $J^{IJ}=(J^{ij}$, $J^{\ipr\jpr})$
are given by
\be\label{kinij2} J^{IJ} = J_0^{IJ} + \sum_{\cI=1,2} \sum_{n=1}^\infty
( a_n^{\cI I} \bar{a}_n^{\cI J} - a_n^{\cI
J}\bar{a}_n^{\cI I}  + \frac{1}{2}\eta_n^\cI
\bar{\gamma}^-\gamma^{IJ}\bar{\eta}_n^\cI )\ ,  \ee
where $J_0^{IJ}$
is the  contribution of the zero modes
\be\la{jjj}
 J_0^{IJ} = a_0^I \bar{a}_0^J-a_0^J\bar{a}_0^I
+\frac{1}{2}\sum_{\cI=1,2} \theta_0^\cI\bar{\gamma}^-\gamma^{IJ}\theta_0^\cI \ . \ee
Note that the kinematical
 generators  do not involve the matrix  $\Pi$ and formally look as if
the  $SO(8)$ symmetry were present.

The dynamical supercharges  \rf{dyncha3} have the following
explicit form \be\la{qwe} \sqrt{p^+} Q^{-1}
=2p_0^I\bar{\gamma}^I\theta_0^1-2\mm
x_0^I\bar{\gamma}^I\Pi\theta_0^2 + \sum_{n=1}^\infty
(2\sqrt{\omega_n}c_n a_n^{1I}\bar{\gamma}^I\bar{\eta}_n^1
+\frac{{\rm i}\mm}{\sqrt{\omega_n} c_n }a_n^{2 I}\bar{\gamma}^I\Pi
\bar{\eta}_n^2 + h.c.)\ee \be\la{qwee} \sqrt{p^+}Q^{-2}
=2p_0^I\bar{\gamma}^I\theta_0^2 + 2\mm
x_0^I\bar{\gamma}^I\Pi\theta_0^1 + \sum_{n=1}^\infty(
2\sqrt{\omega_n}c_n a_n^{2I}\bar{\gamma}^I\bar{\eta}_n^2
-\frac{{\rm i}\mm} {\sqrt{\omega_n} c_n }a_n^{1
I}\bar{\gamma}^I\Pi \bar{\eta}_n^1 + h.c.)\ee These expressions
explicitly  break the $SO(8)$ invariance  down to $SO(4) \times
SO'(4)$.

The requirement that the light-cone gauge formulation respects
basic global symmetries amounts to the condition that the above
generators satisfy the  relations of the symmetry superalgebra of
the plane wave R-R background. The commutators of the bosonic
generators are\foot{ Note that we use the Hermitean $P^\mu$ and
the antiHermitean $J^{\mu\nu}$ generators. The supercharges
$Q^\pm$ and  $\bar{Q}^\pm$ are related to each other by the
conjugation $(Q^\pm)^\dagger =\bar{Q}^\pm$.}
\be \la{pminpi}  [P^-,P^I] = \f^2
J^{+I}\,,\ \ \ \ \ \ \ \ \ \
 [P^I,J^{+J}] = -\delta^{IJ}P^+\,, \qquad
[P^-,J^{+I}]=P^I\,,\ee
\be [P^i,J^{jk}] = \delta^{ij}P^k -\delta^{ik}P^j\,, \qquad
[P^\ipr,J^{\jpr\kpr}] = \delta^{\ipr\jpr}P^\kpr
-\delta^{\ipr\kpr}P^\jpr\,,\ee
\be [J^{+i},J^{jk}] = \delta^{ij}J^{+k} -\delta^{ik}J^{+j}\,,
\qquad [J^{+\ipr},J^{\jpr\kpr}] = \delta^{\ipr\jpr}J^{+\kpr}
-\delta^{\ipr\kpr}J^{+\jpr}\,,\ee
\be [J^{ij},J^{kl}] = \delta^{jk}J^{il} + 3\hbox{ terms}\,,
\qquad [J^{\ipr\jpr},J^{\kpr\lpr}] = \delta^{\jpr\kpr}J^{\ipr\lpr}
+ 3\hbox{ terms}\,. \ee
The
commutation relations  between the even and odd
generators  are
 \be\label{jq1} [J^{ij},Q_\alpha^\pm] =
 \frac{1}{2}Q_\beta^\pm(\gamma^{ij})^\beta{}_\alpha
\,, \qquad [J^{\ipr\jpr},Q_\alpha^\pm] =
\frac{1}{2}Q_\beta^\pm(\gamma^{\ipr\jpr})^\beta{}_\alpha \,, \ee
 \be [J^{+I},Q_\alpha^-] =
 \frac{1}{2}Q_\beta^+(\gamma^{+I})^\beta{}_\alpha\,, \ee \be
\ \ \ \ \ \ [P^I,Q_\alpha^-] =  \frac{1}{2} \f  Q_\beta^+( \Pi
\gamma^{+I})^\beta{}_\alpha\,, \ \ \ \ \
 [P^-,Q_\alpha^+] =  \f Q_\beta^+\Pi^\beta{}_\alpha\,,
 \ee
together with the commutators that follow from these by complex
conjugation. The anticommutation  relations are

 \be \la{dre}
\{Q_\alpha^+,\bar{Q}_\beta^+\} = 2\gamma^-_{\alpha\beta}P^+\,,\ee
\be \{Q_\alpha^+,\bar{Q}_\beta^-\}
=(\bar{\gamma}^-\gamma^+\bar{\gamma}^I)_{\alpha\beta}P^I
-\f(\bar{\gamma}^-\gamma^+\bar{\gamma}^i\Pi)_{\alpha\beta}J^{+i}
-\f(\bar{\gamma}^-\gamma^+\bar{\gamma}^\ipr\Pi^\prime)_{\alpha\beta}J^{+\ipr}\,,\ee
\be \la{drew} \{Q_\alpha^-,\bar{Q}_\beta^+\}
=(\bar{\gamma}^+\gamma^-\bar{\gamma}^I)_{\alpha\beta}P^I
-\f(\bar{\gamma}^+\gamma^-\bar{\gamma}^i\Pi)_{\alpha\beta}J^{+i}
-\f(\bar{\gamma}^+\gamma^-\bar{\gamma}^\ipr\Pi^\prime)_{\alpha\beta}J^{+\ipr}\,,
\ee \be \label{qminqmin}\{Q_\alpha^-,\bar{Q}_\beta^-\}=
2\gamma^+_{\alpha\beta} P^-
+\f(\bar{\gamma}^+\gamma^{ij}\Pi)_{\alpha\beta}J^{ij}
+\f(\bar{\gamma}^+\gamma^{\ipr\jpr}\Pi^\prime)_{\alpha\beta}J^{\ipr\jpr}\,.
\ee
 One can check directly  that our quantum
generators expressed in terms of the creation/annihilation
operators do  satisfy these (anti)commutations relation.
Note that one recovers the flat-space \lc superalgebra in the limit
$\f \to 0$.
{ As
in the flat superstring  case  the anticommutator
relation  between   the dynamical
generators $Q^-$ and $\bar{Q}^-$ \rf{qminqmin} is valid only on
the physical subspace \rf{cons}.}

\subsection{Choice of fermionic zero-mode vacuum}

The states obtained  by applying the fermionic
zero-mode creation operators  to the  vacuum
form a supermultiplet. States  of that supermultiplet
can be described in different ways depending  on how one picks up
 a   (``Clifford'') vacuum to construct
the tower of other states on top of it.    While it
 is natural to define ``the'' vacuum
  to have zero energy,  this is
not the only possible or necessary  choice
 as we shall discuss below.

In general, the quantum  counterpart of  the zero-mode energy
\rf{eee} may be written as (cf. \rf{zer})

\be E_0 = \f \E_0  \ , \ \ \ \ \ \ \ \ \ \ \ \ \E_0=
a_0^I\bar{a}_0^I -2\theta_0\bar{\gamma}^-\Pi\bar{\theta}_0 + e_0
\ , \ \ \ \ee where ${\theta}_0 = { 1 \ov \sqrt 2} ({\theta}^1_0 +
 i {\theta}^2_0)$ (see \rf{nss})  and
  $e_0$  is a  constant  that should be fixed
from the condition of  the
realization of the superalgebra \rf{dre}--\rf{qminqmin}
at the quantum level.
Note that $E_0=0$ in the flat-space limit $\f \to 0$.

We shall need the following expressions for the zero-mode parts
of some symmetry generators (see \rf{jjj},\rf{qwe},\rf{qwee})
\be  J_0^{IJ} = a_0^I\bar{a}_0^J-a_0^J\bar{a}_0^I +
\bar{\theta}_0\bar{\gamma}^-\gamma^{IJ}\theta_0 \,, \ee \be
 \sqrt{p^+}Q_0^{-} =   2 p_0^I\bar{\gamma}^I\theta_0  + 2{\rm i}
\mm x_0^I\bar{\gamma}^I \Pi\theta_0 \ ,  \ \ \ \ \  \ \ \ \ \ \
\sqrt{p^+} \bar{Q}_0^{-} =  2p_0^I\bar{\gamma}^I\bar{\theta}_0  -
2{\rm i} \mm x_0^I\bar{\gamma}^I \Pi\bar{\theta}_0\,. \ee
Let us introduce instead of $\theta_0$ the following complex
 fermionic zero-mode coordinates
\be \la{thee}\vt_{\R} =\frac{1 +  \Pi}{\sqrt{2}}\theta_0  \ , \ \
\ \ \ \ \ \
 \vt_{\L} =\frac{1 -  \Pi}{\sqrt{2}}\theta_0 \ee
satisfying in view of \rf{ttcom1},\rf{pip} the following relations
\be \{\bar\vt_\R,\vt_\R\}= \frac{1}{4}(1+\Pi)\gamma^+
 \ , \qquad \{\bar{\vt}_\L,\vt_\L\}= \frac{1}{4}(1-\Pi)\gamma^+
\ , \qquad \{\bar\vt_\R,\vt_\L\}= 0\ . \ee In terms of them
\be\label{e01} \E_0 = a_0^I\bar{a}_0^I +
\vt_\L\bar{\gamma}^-\bar{\vt}_\L -
 \vt_\R\bar{\gamma}^-\bar\vt_\R + e_0 \ , \ee
and
 \be Q_0^{-}
=2\sqrt{ \f}\ ( a_0^I\bar{\gamma}^I\vt_\R +
\bar{a}_0^I\bar{\gamma}^I
 \vt_\L)  \,, \ \ \ \ \ \bar{Q}_0^{-} =
2\sqrt{ \f}\ ( \bar{a}_0^I\bar{\gamma}^I\bar\vt_\R +
a_0^I\bar{\gamma}^I \bar\vt_\L)  \,, \ \ \ \ \ \ee \be
 J_0^{IJ} = a_0^I\bar{a}_0^J-a_0^J\bar{a}_0^I +\frac{1}{2}
\bar\vt_\R\bar{\gamma}^-\gamma^{IJ}\vt_\R + \frac{1}{2}
\bar\vt_\L\bar{\gamma}^-\gamma^{IJ}\vt_\L \,  .  \ee Let us now
discuss several   possible definitions of the zero-mode vacuum
(we shall always assume that $\bar{a}_0^I \vac =0$). In  all the
cases below the expression for $J^{IJ}$ will imply that the
vacuum is a scalar with respect to $SO(4) \times SO'(4)$.

First,   we may  define the fermionic zero-mode
vacuum  in the same way
is in the case of the flat space background  by imposing
\be \la{hio}
 \bar \theta_0 \vac =0 \ , \ \ \ \ \ \ \ { \rm i.e. } \ \ \ \ \ \ \
 \bar\vt_\R|0\rangle =0 \,,\qquad \bar\vt_\L|0\rangle
=0 \ . \ee
This is the definition we used in \rf{vacdef}.
 Then
$$  \{ Q_0^-,\bar{Q_0}^-\}|0\rangle = 4\f\{ a_0^I\bar{\gamma}^I\vt_\R
+  \bar{a}_0^I\bar{\gamma}^I
 \vt_\L \,,\, \bar{a}_0^J\bar{\gamma}^J\bar\vt_\R +
a_0^J\bar{\gamma}^J \bar\vt_\L \}\vac $$ \be  =
4\f(\bar{a}_0^J\bar{\gamma}^J \bar\vt_\R)\, (a_0^I\bar{\gamma}^I
 \vt_\R )|0\rangle=
 4\f\bar{\gamma}^I \bar\vt_\R \bar{\gamma}^I
 \vt_\R  |0\rangle
= \f\bar{\gamma}^I (1+\Pi)\gamma^+\bar{\gamma}^I
   |0\rangle= -8\f\bar{\gamma}^+ |0\rangle  \ , \ee
{where we use the relation $\bar{\gamma}^I\Pi\gamma^I=0$}. On
the other hand, from the
 supersymmetry algebra relation  \rf{qminqmin}
 we have
\be \{ Q^-,\bar{Q}^-\}|0\rangle = 2\f\bar{\gamma}^+ P^-|0\rangle
=-2\f\bar{\gamma}^+ e_0 \vac \ , \ee where we used that $
J^{IJ}|0\rangle =0$. Since for the zero modes $E_0 = -P^-$
 we  learn that  here $e_0 = 4$.

Thus the  normal ordering of bosons done in
\rf{ezer} is indeed  consistent with the supersymmetry algebra.
Then  from \rf{e01} we see that acting with $\vt_\L$ ($\vt_\R$) on $\vac$
we increase (decrease)  the energy by one unit.
 The generic   fermionic zero-mode state is
\be\la{opo}
 (\vt_\R)^{n_\R} (\vt_\L)^{n_\L} |0\rangle \ , \qquad
n_\L,n_\R =0,1,2,3,4.\ee
The restriction on the values of  $n_\R$ and
$n_\L$ comes from  $(\vt_\R)^5=0$, \ $(\vt_\L)^5=0$
(the projected fermions have only 4 independent components).
 The corresponding energy spectrum is thus
\be\la{eoo} \E_0(n_\R,n_\L) = 4 - n_\R + n_\L\ .\ee
The values of  the  energy  of the lightest  massless
(type IIB supergravity) string  modes
with no bosonic excitations
thus run from 0 to 8 (in units of $\f$).

The equivalent definition of the vacuum  is obtained by using
the conjugate of \rf{hio}
\be \la{kok} \theta_0 \vac = 0 \ , \ \ \ \ \ {\rm i.e. } \ \ \ \
\vt_\R|0\rangle =0\,,\qquad \ \ \ \ \ \ \vt_\L|0\rangle =0 \ , \ee
so that
\be e_0 = 4\,, \  \ \ \ \
\qquad \E_0(n_\R,n_\L) = 4+ n_\R - n_\L \ . \ee

One may   instead  define the  vacuum  by
\be  \bar\vt_\R|0\rangle =0 \,,\qquad  \ \ \ \ \ \
\vt_\L|0\rangle =0 \ , \ee
leading  to
\be \la{juj}
 e_0 = 8\,, \ \qquad  \ \ \ \ \ \  \ \ \
\E_0(n_\R,n_\L) = 8 - n_\R - n_\L\ , \ee
so that  $\E_0$ again takes  values in the range
 $0,1,\ldots, 8$.

Finally,  another possible choice is \be \vt_\R|0\rangle =0\,, \
\ \ \ \ \ \qquad \bar\vt_\L|0\rangle =0  \ , \ee in which case
one finds that \be \la{best} e_0 = 0\,, \ \ \ \  \ \ \ \ \ \ \ \
\qquad \E_0(n_\R,n_\L) = n_\R + n_\L\,. \ee
 Here also   $\E_0=0,1,\ldots, 8$.
 Note that the  two choices of the vacuum  \rf{hio} and \rf{kok}
preserve the $SO(8)$ symmetry but break the effective 2-d
supersymmetry of the \lc string action \rf{laa}
 (the 2-d vacuum energy does not vanish).
At the same time, the choice \rf{best}  preserves the 2-d
supersymmetry, but breaks the $SO(8)$ symmetry down to $SO(4) \times
SO'(4)$ (cf. \rf{thee}).

All these definitions  of the vacuum
are physically equivalent,
being related by a re-labelling of  the
states in the same ``massless'' supermultiplet.
While  in the last  choice we discussed
 the  vacuum energy constant  $e_0$ is zero
(i.e. the normal ordering constants of the
 bosonic and fermionic zero  modes cancel as they do for the
string oscillation  modes), the advantage of the first definition
we have used above in \rf{vacdef} is that it directly corresponds
to the definition of the fermionic vacuum in   flat space
\ci{GSB,gs0}, i.e. with this definition
 one has a  natural smooth flat space limit.

In the next section we shall determine the spectrum of the type
IIB supergravity fluctuation modes in the  background
\rf{bem},\rf{bemm} and will thus be able to explicitly  interpret
the states \rf{opo} with energies $\E_0=0,1,..., 8$ in terms of
particular supergravity fields.

\newsection{Type IIB  supergravity fluctuation spectrum  \\
in the R-R  \pw background }
The  string states  obtained by acting by the fermionic and bosonic
zero-mode operators  on the vacuum
should be in one-to-one correspondence with the
fluctuation modes of type IIB supergravity fields   expanded near the
\pw background \rf{bem},\rf{bemm}.
Assuming the
choice of the zero-mode vacuum  in \rf{vacdef}
or \rf{hio} and  acting by the products of the fermionic zero-mode operators
one finds the  lowest-lying  states
that can be symbolically represented  as
\be\label{grosta}
\begin{array}{ll} |0\rangle \qquad & \hbox{complex scalar}
\\[7pt]
\theta_0|0\rangle \qquad & \hbox{ spin 1/2 field}
\\[7pt]
\theta_0 \theta_0|0\rangle & \hbox{ complex  2-form  field }
\\[7pt]
\theta_0\theta_0\theta_0|0\rangle & \hbox{ spin 3/2 field}
\\[7pt]
\theta_0\theta_0\theta_0\theta_0 |0\rangle  & \hbox{ graviton and self-dual 4-form  field}
\\[7pt]
..& \hbox{ complex conjugates to the above}
\end{array}
\ee
The
complete  type IIB supergravity spectrum  is
 obtained  by
acting with the  bosonic zero mode creation operators
 $a_0^I$ on the  above  states.

The aim of this  section is to explicitly derive the supergravity spectrum
using  the  standard  field-theoretic  approach,
analogous to the one  used in \ci{kim} for the \adss background.

As a preparation,
it is useful to present  the decomposition
 of the  128+128 physical
transverse  supergravity degrees of freedom  in the \lc gauge
 using the  $SO(8) \to SO(4)\times SO'(4)$ decomposition:\foot{
The number of independent components are indicated in
brackets and $N_{{d.o.f.}}$ is the total number of degrees of
freedom.}
\be {\bf graviton}: \ \ \ \  h_{ij}^\perp(9)\,,\quad
h_{\ipr\jpr}^\perp(9)\,,\quad h_{i\jpr}(16) \ ,  \quad h(1) \ ; \
\ \qquad \ \ N_{{d.o.f.}}= 35 \   \ee
 ($h_{ij}^\perp$,
$h_{\ipr\jpr}^\perp$ are traceless and the $h_{i\jpr}$ is not
symmetric in $i,\jpr$)
\be {\bf 4-form \ field}: \ \ \ \ \
a_{i\jpr}(16)\,,\quad a_{ij\ipr\jpr}(18)\,,\quad a(1) \ ;  \
\qquad \ \  \ \ \ N_{{d.o.f.}} =35 \   \ee
($a_{i\jpr}$ is
not antisymmetric in  $i,\jpr$ and  $ a_{ij\ipr\jpr}
=-\frac{1}{4}\epsilon_{ijkl}\epsilon_{\ipr\jpr\kpr\lpr}a_{kl\kpr\lpr}
$) \be {\bf complex\ 2-form\ field}:   \ \
 b_{ij}(12)\,,\quad b_{\ipr\jpr}(12)\,, \quad
b_{i\jpr}(32) \ ; \quad N_{{d.o.f.}}=  56  \  \ee
($b_{i\jpr}$ is not antisymmetric in  $i,\jpr$)
 \be {\bf
complex\ scalar\ field}: \qquad\qquad\ \
 \phi(2) \ ;  \qquad\qquad\qquad \qquad \ \ \ N_{{d.o.f.}} = 2  \ . \ee
\be {\bf spin\ 1/2\ field}:
 \qquad\qquad \qquad
\lambda^\oplus(16)\,;\qquad\qquad\qquad\qquad \ \ \ \ \ \
 N_{d.o.f} = 16 \
\ee
($\l$ is negative chirality complex spinor, and
$\lambda^\oplus = \frac{1}{2}\bar{\gamma}^-\gamma^+\lambda$ is
 its  \lc projection)
\be {\bf spin\ 3/2\  field}:
 \qquad
 \psi_i^{\oplus\perp}(48)\,,\quad
\psi_\ipr^{\oplus\perp}(48)\,,\quad \psi^{\oplus\Vert}(16)\,;\qquad
\ N_{d.o.f}=112  \ee
 (the gravitino is a
positive chirality complex spinor, and  $\psi^\perp$
and  $\psi^\Vert$ are  its $\gamma$-transverse and
 $\gamma$-parallel parts).

As we have  already   found in   string theory  (and
will confirm
  directly from the supergravity equations below),
 here, as in the case of  the AdS supermultiplets,
  the spectrum of  the lowest
eigenvalues of the \lc  energy operator is non-degenerate, i.e.
different states  have   different values of $E_0$.

\subsection{Massless field equations in \pw geometry}

Our aim will be to find the explicit form of the type IIB
equations of motion expanded  to linear order in fluctuations
near the \pw  background  \rf{bem},\rf{bemm} and then to determine
the corresponding \lc energy
spectrum.
 Let us first  discuss the solutions of the simplest wave equations
in the curved metric \rf{bem}.
The non-trivial
components of the corresponding
connection and  curvature  are
 ($g^{--}= \f^2x_I^2$): \be
\Gamma^\mun_{+I} = -\f^2x^I\delta_-^\mun \,,\qquad
\Gamma_{++}^\mun = \f^2x^I\delta_I^\mun\ ,  \ \ \ \ \
R_{I++J}=-\f^2\delta_{IJ}\,,\qquad R_{++}=8\f^2 \ . \ee
The  massless scalar  equation  in the \pw geometry has the
following explicit form
\be \la{sca}  \Box \vp =0 \ , \ \ \ \  \ \ \ \
\Box \equiv  { 1 \ov \sqrt { - g} }
 \del_\mun ( \sqrt { - g}  g^{\mun\nun} \del_\nun)  =
2\partial^+\partial^-
+ \f^2x_I^2\partial^{+2}+\partial_I^2 \ . \ee
After the Fourier transform  in $x^-,x^I$
 corresponding to  the \lc description where
$x^+$ is the evolution parameter
\be \vp(x^+,x^-,x^I) =\int
\frac{dp^+d^{8}p}{(2\pi)^{9/2}}\ e^{{\rm i }(p^+x^-
+p^Ix^I)}\ \tilde \vp(x^+,p^+,p^I) \  \ee
it becomes
\be\la{scaa} (2p^+P^- -\f^2 p^{+2}\partial_{p^I}^2 + p_I^2)\tilde \vp=0\ , \ee
where
$-P^-={\rm i}\partial^-$  may thus be interpreted as   the \lc
Hamiltonian appearing in the non-relativistic Schrodinger equation
for the free harmonic oscillator in 8 dimensions
with mass $p^+$ and frequency  $\f$:
\be \la{duu} H=-P^-  =
 \frac{1}{2p^+}(p_I^2 - \mm^2   \partial_{p^I}^2 ) \ , \ \ \ \ \ \
\ \mm \equiv  \f p^+ \ .  \ee
Introducing the  standard  creation and annihilation operators
\be a^I \equiv \frac{1}{\sqrt{2\mm}}(p^I -
{\mm}\partial_{p^I})\ , \ \ \ \ \ \
 \bar{a}^I \equiv
\frac{1}{\sqrt{2\mm}}(p^I +
{\mm}\partial_{p^I})\ , \ \ \ \
 [\bar{a}^I,a^J]=\delta^{IJ}\ , \ee
we get the following  normal-ordered form of the Hamiltonian
\be H = \frac{1}{2}\f( \bar a^I a^I   +  a^I\bar a^I) =
 \f( a^I\bar a^I + 4)  \ , \ee
where $ 4 = \frac{D-2}{2} , \ D=10$.
As usual,  the spectrum  of states (and thus the solution of \rf{sca})
 is then found by acting by $a^I$ on the vacuum  satisfying
$\bar a^I \vac =0$.

Below  we will need  the following simple generalization of this
analysis:
if  a field $\vp$ satisfies the following
equation
\be\la{ccc}
 (\Box + 2{\rm i}\f c \partial^+)\vp(x) =0 \ ,  \ee
where $\Box$ is defined in \rf{sca}  and $c$ is an arbitrary
 constant,
then the corresponding \lc  Hamiltonian is
\be H=- P^- = \frac{p_I^2- \f^2 p^{+2}\partial_{p^I}^2}{2p^+} + \f c
=  \f( a^I\bar a^I + 4 + c )
 \ , \ee
so that the    lowest \lc energy value is given by
 \be\la{low}
  E_0 = \f \E_0  \ , \ \ \ \ \ \ \ \ \ \ \E_0=4+c \ . \ee
In what follows we shall
 discuss  in turn  the equations of motion for various fields of
type IIB supergravity reducing them to the form
\rf{ccc} and thus determining the corresponding
 lowest energy values from \rf{low}.

\subsection{Bosonic fields }

\noindent
{\bf Complex scalar field}

The dilaton and R-R scalar are decoupled  from the 5-form background
\rf{bemm}, i.e. satisfy
\be \Box\phi =0\ , \ \ \ \ \ \ \ { \rm i.e.} \ \ \ \ \ \  \ \ \E_0(\phi)=4 \ .  \ee

\bigskip
\noindent
{\bf Complex  2-form field}

The corresponding nonlinear equations   are \ci{schwarz}
\be\label{schequ} D^\mun G_{\mun\, \mun_1\mun_2} = P^\mun
G^*_{\mun\,\mun_1\mun_2} - \frac{\rm i }{3} F_{\mun_1\ldots
\mun_5}G^{\mun_3\mun_4\mun_5} \ee where $ G_{\mun_1\mun_2\mun_3}
= 3\partial_{[\mun_1} B_{\mun_2\mun_3]}$ is the field strength of
the complex 2-form field $B_{\mun\nun}$ and $P_\mun $  is the
complex scalar field strength. The  aim is to derive the
equation  for small fluctuations $B_{\mun\nun}=b_{\mun\nun}$ in
the  \pw  background \rf{bem},\rf{bemm} (with  $P_\mun=0 $)
using  the  light-cone gauge \be \la{blc} b_{-\mun}=0  \ .  \ee
It is sufficient to analyze the equations \rf{schequ} for the
following values of the indices $(\mun_1,\mun_2)$: $(-,I)$  and
$(I,J)$. We  find \be \label{IJequ} D^\mun G_{\mun\, IJ } =
\partial_\mu G_{\mu IJ}+\f^2x_I^2\partial^+G_{-IJ} \ , \  \  \ \
\ \ D^\mun G_{\mun\, - I } = \partial_\mu G_{\mu - I}\ .  \ee
Taking into account that $F_{-\mun_2\ldots\mun_5}=0$ and the
light-cone gauge  condition \rf{blc} we find \be \partial^+
b_{+I} +\partial^J b_{JI}=0\,,\ee which allows us to express the
non-dynamical  modes $b_{+I}$ in terms of the physical ones
$b_{IJ}$.
 Then
\be\label{IJequ2}  D^\mun G_{\mun\, IJ } = \Box b_{IJ}\,. \ee
Using that $F_{i\jpr
\mun_3\mun_4\mun_5}=0$ (cf. \rf{bemm})  and
$F_{ij\mun_3\mun_4\mun_5}G^{\mun_3\mun_4\mun_5} =
6\f \epsilon_{ijkl}\partial^+ b_{kl}$  we get from
\rf{schequ},\rf{IJequ2} the following equations  for the
physical modes $b_{IJ}$
\be \Box b_{i\jpr}=0\ , \ \ \ \ \ \  \Box b_{ij} +2{\rm i} \f\epsilon_{ijkl}\partial^+b_{kl}=0 \ , \ \ \ \
 \Box b_{\ipr\jpr} +2{\rm i}  \f\epsilon_{\ipr\jpr\kpr\lpr}
\partial^+ b_{\kpr\lpr}=0\,.\ee
The equation
for $b_{i\jpr}$ implies   that $\E_0(b_{i\jpr})=4$ (see \rf{sca},\rf{low}). To
diagonalize  the remaining equations we
decompose the antisymmetric tensor field
$b_{ij}$ into the irreducible
tensors of the $so(4)$ algebra
\be b_{ij} = b_{ij}^\oplus + b_{ij}^\ominus\,,\qquad \ \ \ \
b_{ij}^{\oplus,\ominus} = \pm
\frac{1}{2}\epsilon_{ijkl}b_{kl}^{\oplus,\ominus} \ . \ee
   Then
\be (\Box + 4{\rm i} \f\partial^+) b_{ij}^\oplus=0 \ , \ \ \ \ \ \ \ \ \
 (\Box - 4{\rm i} \f  \partial^+) b_{ij}^\ominus=0 \ .  \ee
The same relations are found for  $ b_{\ipr\jpr}$.  Then
according to \rf{ccc},\rf{low}
 we  find  the following lowest energy  values
\be \E_0(b_{ij}^\ominus)= 2\,,\qquad \E_0(b_{ij}^\oplus)=
6\,,\qquad \E_0(b_{\ipr\jpr}^\ominus)= 2\,,\qquad
\E_0(b_{\ipr\jpr}^\oplus)= 6\,,\qquad \E_0(b_{i\jpr})=4 \ .  \ee
In the oscillator construction of section  2.4 (see
\rf{hio},\rf{opo},\rf{eoo}) the monomials of the second order in
$\theta_{\L,\R}$ with $\E_0=4$ are $\theta_\R\theta_\L$, which
have $ 16$ complex components,  i.e. these  monomials can be
identified with the  ground state of $b_{i\jpr}$.  The 2-nd
 and 6-th order monomials in  $\theta_\R$,
$\theta_\L$ which can be identified with the ground states of
$b_{ij}^\ominus$, $b_{\ipr\jpr}^\ominus$, $b_{ij}^\oplus$,
$b_{\ipr\jpr}^\oplus$ may be found in Table I.

\bigskip

\noindent
{\bf Graviton and  4-form field}

 Since both the graviton and  the 4-form field have non-trivial backgrounds,  some of their  fluctuation modes are
mixed and need to be analyzed  together. The  full non-linear
form of the  corresponding
 equations of motion are\foot{The equation $DF_5=0$
follows  of course from the self-duality of $F_5$, but we will
find it useful to use this 2-nd order form  of the
equation for $A_4$ below.  Note that we ignore the  quadratic
2-form correction term in $F_5$ \ci{schwarz} as it does not
contribute to the linear fluctuation equations here.}
\be\label{rmn0} R_{\mun\,\nun}=\frac{1}{24}F_{\mun\,\mun_2\ldots
\mun_5}F_\nun{}^{\mun_2\ldots \mun_5} \ee \be\label{self1}
F_{\mun_1\ldots \mun_5 }
=-\frac{1}{5!}\sqrt{-g}\epsilon_{\mun_1\ldots
\mun_5\nun_1\ldots\nun_5} F^{\nun_1\ldots \nun_5} \ , \ \ee \be
\la{chto}
  D^\mun F_{\mun\,\mun_2\ldots \mun_5}=0 \ ,  \ \ \ \ \ \ \
\ \ \ \ \ F_{\mun_1\ldots \mun_5}= 5\partial_{[\mun_1}
A_{\mun_2\ldots \mun_4]}\ . \ee Expanding near the { plane wave
R-R} background \be g_{\mun\,\nun}\to  {
g}_{\mun\,\nun}+h_{\mun\,\nun}\,,\qquad A_{\mun_1\ldots \mun_4}
\to  { A}_{\mun_1\ldots \mun_4} +a_{\mun_1\ldots\mun_4} \ , \ee
\be \ R_{\mun\,\nun} \to  { R}_{\mun\,\nun} +
r_{\mun\,\nun}\,,\qquad \ \ \ F_{\mun_1 \ldots \mun_5}\to  {
F}_{\mun_1 \ldots \mun_5} + f_{\mun_1 \ldots \mun_5} \ , \ee
 we shall choose the \lc gauges for the
fluctuations $h_{\mun\nun}$ and $a_{\mun_1\ldots \mun_4}$
\be\la{kkk}
 h_{-\mun}=0\,, \ \ \ \ \qquad a_{-\mun_2\mun_3 \mun_4}=0\,.\ee
The linearized form of the Einstein equation  is
\be\label{rmn} r_{\mun\,\nun}=\frac{1}{24}(
{ F}_{\mun\,\mun_1\ldots \mun_4}f_\nun{}^{\mun_1\ldots \mun_4}
+ f_{\mun\,\mun_1\ldots \mun_4}{ F}_\nun{}^{\mun_1\ldots
\mun_4} - 4{ F}_{\mun\,\nun_1\mun_3\mun_4
\mun_5}{ F}_{\nun\nun_2}{}^{\mun_3\mun_4\mun_5}
h^{\nun_1\nun_2})
\ee  where
\begin{eqnarray}
r_{\mun\,\nun} &=& \frac{1}{2}\Bigl(-D^2h_{\mun\,\nun}
+D_\mun D^\kun h_{\kun\nun}
 +D_\nun D^\kun h_{\kun\mun}   -D_\mun D_\nun  h_\kun^\kun
\nonumber\\
& +&  2{R}_{\mun\,\mun_1\mun_2\nun}h^{\mun_1\mun_2}    + {R}_{\mun\,\kun}h_\nun^\kun
+{R}_{\nun\,\kun}h_\mun^\kun\Bigr)\,.
\end{eqnarray}
The  $(--)$ component  gives
 $r_{--}=0$  and  thus we find the zero-trace
 condition for the transverse
modes of the graviton \be  h_{II}=0\,. \ee The  $(-I)$ components
of \rf{rmn} gives
 $r_{-I}=0$ and this leads to the equation $D^\mun h_{\mun I}=0$
which allows us to express the  non-dynamical
 modes in terms of the
 physical modes represented by the traceless  tensor
 $h_{IJ}$
\be h_{+I} =-\frac{1}{\partial^+}
\partial_J h_{J I}\,.
\ee
Next, we need to consider the  self-duality
 equation for the 5-form field
whose  $(I_1I_2I_3I_4-)$  component implies that
 $a_{+I_1I_2I_3}$ is expressed  in terms of
the physical modes $a_{IJKL}$
\be
a_{+I_1I_2I_3}^{\phantom{\oplus}}=-\frac{1}{\partial^+}
\partial_J^{\phantom{\oplus}}
a_{JI_1I_2I_3}^{\phantom{\oplus}}\,.   \ee
In terms of  $a_{IJKL}$
 the 5-form field strength self-duality condition  becomes
\be\label{self2} a_{I_1\ldots I_4}^{\phantom{\oplus}} =
-\frac{1}{4!} \epsilon_{I_1\ldots I_4J_1\ldots
J_4}^{\phantom{\oplus}} a_{J_1\ldots
J_4}^{\phantom{\oplus}}\,.\ee
The $(++)$ component of \rf{rmn}
leads to  the expression for $h_{++}$
(after taking into account the above results):
$ h_{++} = {1 \ov (\del^+)^2} \del_I \del_J h_{IJ} $.
So far all is just as in the  \lc analysis near flat space.

Let us now do the $4+4$ split of the 8 transverse directions.
   The  $(i,j)$  components
 of  \rf{rmn}  take the form
\be r_{ij} = \f \delta_{ij}\partial^+ a\,,\qquad \ \ \ \ \ a\equiv \frac{1}{6}\epsilon_{i_1\ldots i_4} a_{i_1\ldots i_4}\, . \ee
Using that $r_{ij} = -\frac{1}{2}\Box h_{ij}$  we get
\be \Box h_{ij} + 2 \f \delta_{ij}\partial^+ a=0\,. \ee
Thus
there is  a mixing  between the  trace of the $SO(4)$ part of the
graviton $h_{ii}$ and the (pseudo) scalar part of the
4-form potential.
 From the $(i_1i_2i_3i_4)$ component of the
 $DF=0$   equation  for the  4-form  field  in \rf{chto}
 we also find
that
\be \label{mix2} \Box a -8\f\partial^+ h_{ii} =0\,.\ee
These equations are diagonalized by introducing
the traceless graviton and the  complex scalar
\be \la{uio}
 h_{ij}^\perp \equiv h_{ij}-\frac{1}{4}\delta_{ij} h_{kk}\,,
\ \ \ \ \ \
\H\equiv h_{ii}+{\rm i} a \,,\qquad \   \bar{\H}\equiv
h_{ii} - {\rm i} a \,,\ee
so that  we finish with
\be\la{fii}
 \Box h_{ij}^\perp =0\,,\ \ \ \ \ \ \  \ \ \
 (\Box -8{\rm i}\f\partial^+)\H=0\,,\qquad (\Box + 8{\rm
i}\f\partial^+)\bar{\H}=0\,.\ee According to  \rf{low} this
implies \be  \E_0( h_{ij}^\perp) = 4 \ , \ \ \ \ \ \ \
 \E_0(\H)= 0\,, \ \  \qquad \E_0(\bar{\H}) =8\,. \ee
The same results are found of course in the other four   directions,
i.e. with $h_{ij} \to h_{\ipr\jpr}$ and $a \to a'= \frac{1}{6}
\epsilon_{\ipr_1\ldots \ipr_4} a_{\ipr_1\ldots \ipr_4}$,  \ $a'=-a$.

Let us now look at ``mixed'' components. Eqs. \rf{rmn}
in $(i\jpr)$ directions give
\be\label{mix3} \Box h_{i\jpr} +
4\f\partial^+a_{i\jpr}=0\,,  \ \ \ \ \  \ \ \ \
a_{i\jpr}\equiv  {1 \ov 3}
\epsilon_{ii_1i_2i_3}a_{\jpr i_1i_2i_3}\ .
\ee
 We have used the  self-duality
 \rf{self2} implying
$ \epsilon_{ii_2i_3i_4} a_{\jpr i_2i_3i_4}= \epsilon_{\jpr
\ipr_2\ipr_3\ipr_4} a_{i \ipr_2\ipr_3\ipr_4}$.
In addition, the $(i\jpr j_1'j_2')$
 components  of the $DF=0$ equations \rf{chto}
  give
\be\label{mix4} \Box a_{i\jpr} -4\f\partial^+h_{i\jpr}=0\,.\ee
Again   there is  a
 mixing between the components of the
 graviton and the  4-form  field.
These equations are diagonalized by defining
the  complex tensor
\be \H_{i\jpr} \equiv h_{i\jpr} + {\rm i} a_{i\jpr}\,,
\qquad \ \ \ \
\bar{\H}_{i\jpr} \equiv h_{i\jpr} -{\rm i}
a_{i\jpr} \ , \ee
\be (\Box -4{\rm i}\f\partial^+ )\H_{i\jpr}=0\,,\qquad  (\Box
+4{\rm i}\f\partial^+ )\bar{\H}_{i\jpr}=0\,,\ee
so that the  corresponding  lowest eigenvalues of the
energy are
\be \E_0(\H_{i\jpr})= 2\,, \qquad\ \ \ \
\  \E_0(\bar{\H}_{i\jpr})= 6\,. \ee
Finally,  for  $a_{ij\ipr\jpr}$  satisfying, according
 to
\rf{self2}, the constraint
\be a_{ij\ipr\jpr}^{\phantom{\oplus}} =
-\frac{1}{4}\epsilon_{ijkl}^{\phantom{\oplus}}
\epsilon_{\ipr\jpr\kpr\lpr}^{\phantom{\oplus}}
a_{kl\kpr\lpr}^{\phantom{\oplus}}\,\ee
 we find from \rf{chto}  that
\be \Box a_{ij\ipr\jpr}=0 \,,\ \ \ \ \ {\rm  i.e.} \ \ \ \ \ \
\E_0(a_{ij\ipr\jpr})=4\,.\ee Note that the self-dual tensor field
$a_{ij\ipr\jpr}$ is reducible with respect to the $SO(4)\times
SO'(4)$ group. It can be decomposed into the irreducible parts
$a_{ij\ipr\jpr}^{\oplus\ominus}$, $a_{ij\ipr\jpr}^{\ominus\oplus}$
satisfying
\be a_{ij\ipr\jpr}^{\oplus\ominus}=
\frac{1}{2}\epsilon_{ijkl}^{\phantom{\oplus}}a^{\oplus\ominus}_{kl\ipr\jpr}\,,\qquad
a_{ij\ipr\jpr}^{\oplus\ominus}=
-\frac{1}{2}\epsilon_{\ipr\jpr\kpr\lpr}^{\phantom{\oplus}}
a^{\oplus\ominus}_{ij\kpr\lpr}\,, \ee
\be a_{ij\ipr\jpr}^{\ominus \oplus}=
-\frac{1}{2}\epsilon_{ijkl}^{\phantom{\oplus}}a^{\ominus\oplus}_{kl\ipr\jpr}\,,\qquad
a_{ij\ipr\jpr}^{\ominus\oplus}=
\frac{1}{2}\epsilon_{\ipr\jpr\kpr\lpr}^{\phantom{\oplus}}
a^{\ominus\oplus}_{ij\kpr\lpr}\,. \ee
The $SO(4)\times SO'(4)$ labels of these
irreducible parts may be found in Table 1.

\subsection{Fermionic  fields }

Let us now extend the above analysis  to the
fermionic fields  of type IIB supergravity.

\bigskip
\noindent
{\bf Spin 1/2 field}

 The equation of motion for the  two Majorana-Weyl
 negative
chirality  spin 1/2  fields combined into one $32$-component
Weyl spinor field $\Lambda$  \ci{schwarz}

\be (\Gamma^\mun D_\mun -\frac{\rm i}{480}\Gamma^{\mun_1\ldots
\mun_5}F_{\mun_1\ldots \mun_5})\Lambda=0\ , \ee
can be rewritten in terms of the
 complex-valued
$16$-component spinor field $\lambda$ (see Appendix for notation)
\be (\gamma^\mun D_\mun - \frac{\rm i}{480}\gamma^{\mun_1\ldots
\mun_5}F_{\mun_1\ldots \mun_5})\lambda=0\ , \ \ \ \ \ \ \ \
\Lambda =\left(\begin{array}{c} 0 \\[7pt]
\lambda_\alpha\end{array}\right) \ .
\ee
Here $ \gamma^\mun = e^\mun_\mu \g^\mu$ where $e^\mun_\mu$ is the (inverse) vielbein matrix.
We use the following vielbein basis
corresponding to the metric \rf{bem} ($e^\mu = e^\mu_\mun dx^\mun$)
\be e^+ = dx^+ \,,\qquad e^- = dx^- -\frac{\f^2}{2}x_I^2
dx^+\,,\qquad e^I = dx^I\,.\ee
 The  spinor covariant
derivative $  D_\mun =\partial_\mun
+\frac{1}{4}\omega_\mun^{\,\,\mu\nu}\bar{\gamma}^{\mu\nu} $
then takes  the following explicit form
 \be D_- = \partial_- \ , \ \ \qquad D_I =\partial_I\,,\
\ \qquad  D_+ =\partial_+ -\frac{\f^2}{2}x^I\bar{\gamma}^{+I} \ .
\ee
 Taking into account  the background value
of the 5-form field \rf{bemm} we get \be \Bigl[\gamma^+(\partial^-
+\frac{\f^2}{2}x_I^2\partial^+ - {\rm i} \f\bar{\Pi})
+\gamma^-\partial^+ + \gamma^I\partial^I\Bigr]\lambda =0\,, \ee
where  we used that \be \la{gamf} \gamma^{\mun_1\ldots \mun_5}
F_{\mun_1\ldots \mun_5} = 480 \f \gamma^+\bar{\Pi}  \ . \ee
 Decomposing
 $\lambda$ as
\be\label{dec9}
 \lambda = \lambda^\oplus + \lambda^\ominus \  , \ \ \ \ \
 \lambda^\oplus =
\frac{1}{2}\bar{\gamma}^-\gamma^+\lambda\,, \ \
\quad \lambda^\ominus
= \frac{1}{2}\bar{\gamma}^+\gamma^-\lambda\ ,  \ee
we find that in the \lc description
$\lambda^\ominus$ is  non-dynamical mode expressed
in terms of the physical mode $\lambda^\oplus$
\be \lambda^\ominus =
\frac{1}{2\partial^+}\bar{\gamma}^I\partial^I\gamma^+
\lambda^\oplus\,, \ \ \ \ \ \ \
 (\Box - 2{\rm i}\f \bar{\Pi}\partial^+)\lambda^\oplus =0\,.\ee
Decomposing $\lambda^\oplus$ further as (cf. \rf{thee})
\be \label{pidec}\lambda^\oplus =\lambda^\oplus_\R +
\lambda_\L^\oplus\,,  \ \ \ \qquad \lambda_\R\equiv
\frac{1+\bar{\Pi}}{2}\lambda\,, \ \ \ \ \qquad
 \lambda_\L\equiv \frac{1-\bar{\Pi}}{2}\lambda\,, \ee
we  get the diagonal equations of  the desired form \rf{ccc} \be
(\Box  - 2{\rm i}\f \partial^+)\lambda_\R^\oplus =0\,,\ \ \ \ \ \
(\Box + 2{\rm i}\f \partial^+)\lambda_\L^\oplus =0\,,\ee Then from
\rf{low} we conclude that
 the lowest values of the \lc energy for the
fields  $\lambda_\R^\oplus$, $\lambda_\L^\oplus$ are \be
\E_0(\lambda_\R^\oplus) =3\,,\qquad\ \ \ \ \
  \E_0(\lambda_\L^\oplus)
=5\,. \ee

\bigskip

\noindent
{\bf Spin 3/2 field}

 The equation  for the positive chirality
 gravitino
 in the $32$-component notation is\foot{The 5-form term
in the gravitino equation was missing in \ci{schwarz}
but its presence is
 implied by the supersymmetry transformations  given there
and in \ci{west}. This term was explicitly included in \ci{kim}.}

\be \Gamma^{\mun\,\mun_1\mun_2}\Bigl(D_{\mun_1} +\frac{\rm
i}{960}\Gamma^{\nun_1\ldots \nun_5}F_{\nun_1\ldots
\nun_5}\Gamma_{\mun_1}\Bigr)\Psi_{\mun_2}=0\,.\ee
In the
 $16$-component notation it becomes
\be \bar{\gamma}^{\mun\,\mun_1\mun_2}\Bigl(D_{\mun_1} +\frac{\rm
i}{960}\gamma^{\nun_1\ldots \nun_5}F_{\nun_1\ldots
\nun_5}\bar{\gamma}_{\mun_1}\Bigr)\psi_{\mun_2}=0\,,
\ \ \ \ \ \ \ \
 \Psi_\mun =\left(\begin{array}{c} \psi_\mun^\alpha \\[7pt]
0 \end{array}\right)\,.\ee
 This can be rewritten as
\be\label{gra3} \bar{\gamma}^\nun D_\nun \psi_\mun -D_\mun \rhoh
-\frac{\rm i}{960}\bar{\gamma}^\nun \gamma^{\nun_1\ldots
\nun_5}F_{\nun_1\ldots \nun_5}\bar{\gamma}_\mun \psi_\nun
=0\,,\qquad \ \ \  \rhoh \equiv \bar{\gamma}^\mun\psi_\mun\,. \ee  Making use of  \rf{gamf}
we get
\be\label{gra4} \bar{\gamma}^\nun D_\nun \psi_\mun
-D_\mun\psi-\frac{{\rm i\f}}{2} \bar{\gamma}^\nun
\Pi\gamma^+\bar{\gamma}_\mun \psi_\nun =0\,,\ee and
impose  the light-cone gauge for the gravitino field
\be \psi_-=0\,.\ee
 Eq. \rf{gra4} for $\mun=-$ then gives
\be \label{rhosol}  \rhoh=\bar{\gamma}^+\psi_+ +
\bar{\gamma}^I\psi_I=0\ , \ \ \ \quad {\rm i.e.} \ \ \
\quad\gamma^+\bar{\gamma}^I\psi_I=0\,. \ee
As a consequence,
\be \bar{\gamma}^J\Pi \gamma^+\bar{\gamma}_i\psi_J
=2\bar{\Pi}\bar{\gamma}^+(\delta_{ij}-\gamma_i\bar
{\gamma}_j)\psi_j\ , \ \ \ \ \ \ \
 \bar{\gamma}^J\Pi \gamma^+\bar{\gamma}_\ipr\psi_J= -
2\bar{\Pi}\bar{\gamma}^+(\delta_{\ipr\jpr}
-\gamma_\ipr\bar{\gamma}_\jpr)\psi_\jpr \ . \ee
With the help of these  relations the $\mun=i$ component  of
 \rf{gra4}  becomes
\be\label{gra5}\Bigl[
\bar{\gamma}^+(\partial^-+\frac{\f^2}{2}x_I^2\partial^+)
+\bar{\gamma}^-\partial^+ + \bar{\gamma}^J\partial_J\Bigr]\psi_i
-{\rm
i}\f\bar{\Pi}\bar{\gamma}^+(\delta_{ij}-\gamma_i\bar{\gamma}_j)\psi_j=0\ . \ee
Decomposing the
gravitino field into the  physical mode $\psi_i^\oplus$
and non-dynamical  mode  $\psi_i^\ominus$
 as in \rf{dec9} we get from  eq.
\rf{gra5} (acting by  $\gamma^+$  or  by $\gamma^-$)
\be\label{psi1} \Box \psi_i^\oplus -2{\rm i}\f\
\Pi(\delta_{ij}-\gamma_i\bar{\gamma}_j)\partial^+\psi_j^\oplus=0
\ , \ \ \ \ \ \ \ \ \ \
 \psi_I^\ominus
=-\frac{1}{2\partial^+}\gamma^+
(\bar{\gamma}^J\partial_J)\psi_I^\oplus  \ .
\ee
 The other non-dynamical mode $\psi_+$  (split into
$\psi_+^\oplus$ and $\psi_+^\ominus$ as in \rf{dec9})  is found
from \rf{rhosol} and the $\mun=+$ component of the gravitino
 equation \rf{gra4}
\be \psi_+^\oplus =
-\frac{1}{\partial^+}\partial_I^{\phantom{\oplus}}\psi_I^\oplus\,, \ \ \ \
\qquad \psi_+^\ominus =
-\frac{1}{2\partial^+}\gamma^+\bar{\gamma}^I\partial_I^{\phantom{\oplus}}
\psi_+^\oplus\,. \ee
 Decomposing  the dynamical gravitino mode $\psi_I^\oplus$
into the $\gamma$-transverse and
$\gamma$-parallel parts as
\be\label{psilon1}  \psi_i^{\oplus\perp}\equiv
(\delta_{ij}-\frac{1}{4}\gamma_i\bar{\gamma}_j)\psi_j^\oplus\
 , \ \ \ \ \ \  \ \ \ \ \psi^{\oplus\Vert} \equiv
\bar{\gamma}_i\psi_i^\oplus \ee
 we  find
\be \la{ioi}
(\Box -2{\rm i}\f\Pi\partial^+)\psi_i^{\oplus\perp}=0\,,\ \ \ \ \
\qquad (\Box -6{\rm
i}\f\bar{\Pi}\partial^+)\psi^{\oplus \Vert}=0\,. \ee
 As in the spin 1/2 case,  to
diagonalize these equations we introduce  (cf.  \rf{pidec})
\be\la{pop}
 \psi_{i\R}^{\oplus \perp} =\frac{1+\Pi}{2} \psi_i^{\oplus\perp}\,, \ \ \ 
\psi_{i\L}^{\oplus \perp} =\frac{1-\Pi}{2}\psi_i^{\oplus\perp}\,, \ \ \ 
\psi_\R^{\oplus \Vert} =\frac{1+\bar
\Pi}{2}\psi^{\oplus\Vert}\,,
 \ \ \ 
\psi_\L^{\oplus \Vert} =\frac{1-\bar \Pi}{2}\psi^{\oplus\Vert}\,.
\ee
 This gives finally \be\label{psir} (\Box -2{\rm
i}\f\partial^+)\psi_{i \R}^{\oplus \perp}=0 \ , \ \ 
 (\Box  + 2{\rm i}\f\partial^+)\psi_{i \L}^{\oplus
\perp}=0 \ , 
 \ \ 
 (\Box -6{\rm i}\f\partial^+)\psi_\R^{\oplus \Vert}=0 \ , \
\ 
 (\Box + 6{\rm i}\f\partial^+)\psi_\L^{\oplus \Vert}=0 \ . \ee
These equations give, according to \rf{ccc},\rf{low}
 the following  values of the minimal energy $\E_0$ for  the
respective  physical gravitino  modes \be \E_0(\psi_{i
\R}^{\oplus \perp})=3\,,\ \ \quad \E_0(\psi_{i \L}^{\oplus
\perp})=5\,,\ \ \qquad \E_0(\psi_\R^{\oplus \Vert})=1\,,\ \
\qquad \E_0(\psi_\L^{\oplus \Vert})=7\ . \ee Similar analysis
applies to the  gravitino components $\psi_\ipr$. In this case we
get  (cf.  \rf{psi1})  
\be \Box \psi_\ipr^\oplus + 2{\rm
i}\f\Pi(\delta_{\ipr\jpr}-\gamma_\ipr\bar{\gamma}_\jpr)\partial^+\psi_\jpr^\oplus
= 0 \ , \ee
and as a result
 \be
\E_0(\psi_{\ipr \R}^{\oplus \perp})=5\,,\ \ \ \ \ \ \ \ \ \ \quad
\E_0(\psi_{\ipr \L}^{\oplus \perp})=3\,.\ee As for the
$\gamma$-parallel  part $\psi'{}^{\oplus\Vert} = \bar{\gamma}_\ipr
\psi_\ipr^\oplus$  of $\psi_\ipr$,  it  does not represent an
independent dynamical mode being  related  to
$\psi^{\oplus\Vert}$ through the equation \rf{rhosol},  i.e.  $
\bar{\gamma}^I\psi_I^\oplus =0$.



\subsection{Light-cone gauge superfield formulation of type \\
IIB supergravity on the  plane wave  background }

Before proceeding, let  us first  summarize the results of the above analysis
in the two Tables: one  for the bosonic modes, and another for
 the fermionic modes.

\newpage

TABLE I. {\sf Spectrum of bosonic physical  on-shell  fields}
\begin{center}
\begin{tabular}{|c|c|c|c|c|}
\hline &&&&\\ [-3mm]  & Field & Energy &
 $SO(4) \times SO'(4)$ &  Term in \\
${\cal E}_0$ & \ \ and \ \  & spectrum & labels & superfield expansion \\
 &$N_{d.o.f}$  & k\,$\geq 0$ & &
\\ \hline
&&&&\\[-3mm]
0 & ${\rm h}$\,(2) & k &$(0,0)\times(0,0)$  &  $\theta_\R^4$ \\[2mm]\hline
&&&&\\[-3mm]
2   & ${\rm h}_{i\jpr}$\,(32) & k+2& $(1,0)\times (1,0)$ &
$\theta_\R\bar{\gamma}^{-ik}\theta_\R\theta_\R\bar{\gamma}^{-\jpr
k} \theta_\L $
\\[2mm]\hline
&&&&\\[-3mm]
2   & $\bar{b}_{ij}^\oplus$(6) & k+2& $(1,1)\times (0,0)$ &
$\theta_\R^4 (\theta_\L \prp\bar{\gamma}^{-ij}\theta_\L)$
\\[2mm]\hline
&&&&\\[-3mm]
2   & $\bar{b}_{\ipr\jpr}^\oplus$(6)&  k+2& $(0,0)\times (1,1)$ &
$\theta_\R^4 (\theta_\L \prp\bar{\gamma}^{-\ipr\jpr}\theta_\L)$
\\[2mm]\hline
&&&&\\[-3mm]
2   & $b_{ij}^\ominus$(6)&  k+2& $(1,1)\times (0,0)$ & $
\theta_\R \prm\bar{\gamma}^{-ij}\theta_\R$
\\[2mm]\hline
&&&&\\[-3mm]
2   & $b_{\ipr\jpr}^\ominus$(6)&  k+2& $(0,0)\times (1,-1)$
& $\theta_\R \prm\bar{\gamma}^{-\ipr\jpr}\theta_\R$
\\[2mm]\hline
&&&&\\[-3mm]
4   & $ \phi$\,(2) & k+4&$(0,0)\times (0,0)$ &  $ 1 $
\\[2mm]\hline
&&&&\\[-3mm]
4   & $ \bar{\phi}$\,(2) & k+4&$(0,0)\times (0,0)$ & $\theta_\R^4
\theta_\L^4 $
\\[2mm]\hline
&&&&\\[-3mm]
4   & $ h_{ij}^\perp$(9) & k+4&$(2,0)\times (0,0)$ &
$\theta_\R\bar{\gamma}^{-k(i}\theta_\R\theta_\L\bar{\gamma}^{j)k-}\theta_\L
$
\\[2mm]\hline
&&&&\\[-3mm]
4   & $h_{\ipr\jpr}^\perp$(9) &k+4& $(0,0)\times (2,0)$ &
$\theta_\R\bar{\gamma}^{-\kpr(\ipr}\theta_\R\theta_\L\bar{\gamma}^{\jpr)\kpr-}\theta_\L
$
\\[2mm]\hline
&&&&\\[-3mm]
4   & $a_{ij\ipr\jpr}^{\oplus\ominus}$(9)  &
k+4&$(1,1)\times(1,-1)$ &
 $\theta_\L \prp\bar{\gamma}^{-ij}\theta_\L \theta_\R
\prm\bar{\gamma}^{-\ipr\jpr}\theta_\R$
\\[2mm]\hline
&&&&\\[-3mm]
4   & $a_{ij\ipr\jpr}^{\ominus\oplus}$(9)  & k+4 & $(1,-1)\times
(1,1)$ &  $\theta_\R \prm\bar{\gamma}^{-ij}\theta_\R \theta_\L
\prp\bar{\gamma}^{-\ipr\jpr}\theta_\L$
\\[2mm]\hline
&&&&\\[-3mm]
4   & $b_{i\jpr}$\,(32)  & k+4 & $(1,0)\times (1,0)$ &
$\theta_\R\bar{\gamma}^{-i\jpr} \theta_\L $
\\[2mm]\hline
&&&&\\[-3mm]
6   & $b_{ij}^\oplus$(6) & k+6& $(1,1)\times (0,0)$ &
$\theta_\L\prp\bar{\gamma}^{-ij}\theta_\L $
\\[2mm]\hline
&&&&\\[-3mm]
6   & $b_{\ipr\jpr}^\oplus$(6) &  k+6& $(0,0)\times (1,1)$ &
$\theta_\L\prp\bar{\gamma}^{-\ipr\jpr}\theta_\L $
\\[2mm]\hline
&&&&\\[-3mm]
6   & $\bar{b}_{ij}^\ominus$(6) & k+6& $(1,-1)\times (0,0)$ &
$\theta_\L^4(\theta_\R\prm\bar{\gamma}^{-ij}\theta_\R) $
\\[2mm]\hline
&&&&\\[-3mm]
6   & $\bar{b}_{\ipr\jpr}^\ominus$(6) & k+6&$(0,0)\times (1,-1)$ &
$\theta_\L^4(\theta_\R\prm\bar{\gamma}^{-\ipr\jpr}\theta_\R) $
\\[2mm]\hline
&&&&\\[-3mm]
6   & $\bar{\rm h}_{i\jpr}$\,(32) & k+6 & $(1,0)\times (1,0)$ &
$\theta_\L\bar{\gamma}^{-ik}\theta_\L
\theta_\R\bar{\gamma}^{-\jpr k}\theta_\L$
\\[2mm]\hline
&&&&\\[-3mm]
8& $\bar{\rm h}$\,(2)  & k+8  & $(0,0)\times (0,0)$ &  $\theta_\L^4$\\[2mm]\hline
\end{tabular}
\end{center}

\bigskip
\newpage

TABLE II. {\sf Spectrum of fermionic physical  on-shell  fields}
\begin{center}
\begin{tabular}{|c|c|c|c|c|}
\hline &&&&\\ [-3mm]  & Field & Energy &
 $SO(4) \times SO'(4)$ & Term in  \\
${\cal E}_0$ & \ \ and \ \  & spectrum & labels  & superfield expansion\\
 &$N_{d.o.f}$  & k\,$\geq 0$ & &
\\ \hline
&&&&\\[-3mm]
1 & $\psi_\R^{\oplus\Vert}\,(8) $ & k+1
&$(\half,-\half)\times(\half,\half)$
&  $\theta_\R^3$ \\[2mm]\hline
&&&&\\[-3mm]
1   & $\bar{\psi}_\L^{\oplus\Vert}\,(8)$ & k+1&
$(\half,\half)\times (\half,-\half)$ & $\theta_\R^4\theta_\L$
\\[2mm]\hline
&&&&\\[-3mm]
3   & $\psi_{i\R}^{\oplus\perp}\,(24)$ & k+3&
$(\frac{3}{2},-\half)\times (\half,\half)$ &
$(\theta_\R\bar{\gamma}^{-ij}\theta_\R)\bar{\gamma}^j\theta_\L$
\\[2mm]\hline
&&&&\\[-3mm]
3   & $\psi_{\ipr\L}^{\oplus\perp}\,(24)$ & k+3&
$(\half,\half)\times (\frac{3}{2},-\half)$ &
$(\theta_\R\bar{\gamma}^{-\ipr\jpr}\theta_\R)\bar{\gamma}^\jpr
\theta_\L$
\\[2mm]\hline
&&&&\\[-3mm]
3   & $\bar{\psi}_{i\L}^{\oplus\perp}$\,(24)&  k+3&
$(\frac{3}{2},\half)\times (\half,-\half)$ & $\theta_\R^3
\theta_\L^2$
\\[2mm]\hline
&&&&\\[-3mm]
3   & $\bar{\psi}_{\ipr\R}^{\oplus\perp}$\,(24)&  k+3&
$(\half,-\half)\times (\frac{3}{2},\half)$ &
$\theta_\R^3\theta_\L^2$
\\[2mm]\hline
&&&&\\[-3mm]
3   & $\lambda_\R^\oplus$\,\,(8)&  k+3& $(\half,-\half)\times
(\half,\half)$ & $\theta_\R$
\\[2mm]\hline
&&&&\\[-3mm]
3   & $\bar{\lambda}_\L^\oplus$\,\,(8)&  k+3& $(\half,\half)\times
(\half,-\half)$ & $\theta_\R^4\theta_\L^3$
\\[2mm]\hline
&&&&\\[-3mm]
5   & $\lambda_\L^\oplus$\,\,(8)&  k+5& $(\half,\half)\times
(\half,-\half)$ & $\theta_\L$
\\[2mm]\hline
&&&&\\[-3mm]
5   & $\bar{\lambda}_\R^\oplus$\,\,(8)&  k+5&
$(\half,-\half)\times (\half,\half)$ & $\theta_\R^3\theta_\L^4$
\\[2mm]\hline
&&&&\\[-3mm]
5   & $\psi_{i\L}^{\oplus\perp}$\,(24)&  k+5&
$(\frac{3}{2},\half)\times (\half,-\half)$ &
$(\theta_\L\bar{\gamma}^{-ij}\theta_\L)\bar{\gamma}^j\theta_\R$
\\[2mm]\hline
&&&&\\[-3mm]
5   & $\psi_{\ipr\R}^{\oplus\perp}$\,(24)&  k+5&
$(\half,-\half)\times (\frac{3}{2},\half)$ &
$(\theta_\L\bar{\gamma}^{-\ipr\jpr}\theta_\L)\bar{\gamma}^\jpr
\theta_\R$
\\[2mm]\hline
&&&&\\[-3mm]
5   & $\bar{\psi}_{i\R}^{\oplus\perp}$\,(24)&  k+5&
$(\frac{3}{2},-\half)\times (\half,\half)$ &
$\theta_\R^2\theta_\L^3$
\\[2mm]\hline
&&&&\\[-3mm]
5   & $\bar{\psi}_{\ipr\L}^{\oplus\perp}$\,(24)&  k+5&
$(\frac{3}{2},\half)\times (\half,-\half)$ &
$\theta_\R^2\theta_\L^3$
\\[2mm]\hline
&&&&\\[-3mm]
7   & $\psi_\L^{\oplus\Vert}$\,(8) & k+7& $(\half,\half)\times
(\half,-\half)$ &  $\theta_\L^3 $
\\[2mm]\hline
&&&&\\[-3mm]
7   & $\bar{\psi}_\R^{\oplus\Vert}\,(8)$ & k+7&
$(\half,-\half)\times (\half,\half)$ &  $\theta_\L^4\theta_\R $
\\[2mm]\hline
\end{tabular}
\end{center}

\bigskip

In Tables  I,II in
the  ${\cal E}_0$ column we indicate lowest
eigenvalues of the \lc  energy operator of
the corresponding field.  The energy
spectrum   of  higher ``Kaluza-Klein'' modes
 (obtained  by further action  by the bosonic zero-mode creation
operators $a^I_0$)
is labeled by k, where k$=0$
corresponding  to the ground state.
Note,  however,  that
 these are not the usual  Kaluza-Klein-type
 modes  because
the  action of the symmetry algebra of the
plane wave  background mixes  modes with different values
of k.
This algebra
can be thus viewed as a  spectrum
generating algebra for the  ``Kaluza-Klein'` modes.

In the fourth column we have given the Gelfand-Zetlin
labels of the corresponding $SO(4) \times SO'(4)$ representations.
In the last column  we
indicated  the
monomials in  fermionic zero modes
$\theta_\L, \theta_\R$  which  accompany  the
corresponding  field components in the $\theta$-expansion
 of the light-cone superfield   discussed below.

\bigskip


In the rest of this  section we shall present the
light-cone gauge superfield
description  of type IIB supergravity in the   plane wave R-R
background. As in flat space,
the  equations for the physical modes we have  found
above can be   summarized in a  \lc superfield
form. The corresponding unconstrained scalar superfield  $\Phi(x,\theta_0)$
will
satisfy the ``massless'' equation,
 invariant under the
dilatational invariance in superspace.

Finding even the quadratic part of the action  for fluctuations
of the supergravity fields in a curved background is a
complicated problem.\foot{In the case of the \adss
background in covariant gauge it was solved in
  \ci{far}.}
We could in principle use the covariant
superfield description of type IIB supergavity \ci{HW}, starting
with linearized expansion of superfileds,  imposing light-cone
gauge on fluctuations and then solving the  constraints to
eliminate non-physical degrees of freedom in terms of physical
ones. That would be quite tedious. The  light-cone gauge approach
is self-contained, i.e.  does not rely upon
existence of a covariant description,  and  provides  a much
shorter route to   final  results.

There  are two methods  of finding the  light-cone gauge formulation of
the type II supergravity. One  \ci{dir} reduces the problem
of constructing a new (light-cone gauge) dynamical system  to
 finding a new solution of the commutation relations of
the defining symmetry algebra. This method of Dirac was applied
to the case of supergravity in $AdS_5\times S^5$ and $AdS_3\times
S^3$ in \ci{mt4} and \ci{met3}.\footnote{The application of this
method to a superfield formulation of interaction vertices of
$D=11$ supergravity may be found {in} \ci{met4}(see also
\ci{ple3} for  various related discussions).} The second method
one is based on finding  the  equations of motion by
using the Casimir operators of the symmetry algebra. Here we
shall follow this  second approach.

The basic \lc gauge superfield will be denoted as
$\Phi(x,\theta)$ and will have the following expansion in powers
of the Grassmann coordinates $\theta$ \foot{Here we omit the
index 0 on the \lc fermionic zero-mode variable $\theta_0$
denoting it simply as $\theta$. To simplify the
expressions for the
superfield
expansion  and its reality constraint
 we solve the \lc gauge constraint $\bar \g^+ \theta=0$ in
terms of eight fermions
$\theta^a$ ($a=1,\ldots,8$)
 by using the
representation for $\gamma^0$ in \rf{gammaspl} and
$\gamma^9=\diag(1_8,-1_8)$.
}

\begin{eqnarray}
 \Phi(x,\theta)&=&\partial^{+2} A + \theta^a\partial^+\psi_a
+\theta^{a_1}\theta^{a_2}\partial^+ A_{a_1a_2}
\nonumber\\
&+&\theta^{a_1}\theta^{a_2}\theta^{a_3} \psi_{a_1a_2a_3}
+\theta^{a_1}\ldots\theta^{a_4} A_{a_1\ldots a_4} -
(\epsilon\theta^5)_{a_1a_2a_3} \frac{\rm i
}{\partial^+}\psi^{a_1a_2a_3*}
\nonumber\\
\label{supfield} &-& (\epsilon\theta^6)_{a_1a_2}
\frac{1}{\partial^+}A^{a_1a_2*} +(\epsilon\theta^7)_a\frac{\rm i
}{\partial^{+2}}\psi^{a*}
+(\epsilon\theta^8)\frac{1}{\partial^{+2}} A^*\,,
\end{eqnarray}
where $\epsilon_{a_1\ldots a_8}$ is the spinorial Levi-Civita
tensor, i.e.
\begin{equation}
(\epsilon\theta^{8-n})_{a_1\ldots a_{n}}\equiv\frac{1}{(8-n)!}
\epsilon_{a_1\ldots a_{n}a_{n+1}\ldots a_8}
\theta^{a_{n+1}}\ldots\theta^{a_8} \ .
\end{equation}
Here we  use the following Hermitean conjugation rule: $(\theta_1\theta_2)^\dagger
=\theta_2^\dagger \theta_1^\dagger $.
This superfield  has a certain reality
property:
the component field for
 the monomial $\theta^n$ is complex conjugated to the one
for  $\theta^{8-n}$. This reality constraint can be
written in the superfield notation as

\begin{equation}
\Phi(x,\theta) =\int d^8\theta^\dagger\ e^{{\rm
i}(\partial^+)^{-1} \theta\theta^\dagger} \ (\partial^{+})^4\
(\Phi(x,\theta))^\dagger\,.
\end{equation}
In what follows we will use  again the 16-component
spinor $\theta^\alpha= \left( \begin{array}{c} \theta^a \\
0\end{array}\right)$.
Decomposing  it into $\theta_\R$ and $\theta_\L$ as in
\rf{thee} we can expand the  superfield  $\Phi$ in terms of these
anticommuting coordinates.

The expansion in this basis
 can be used to
identify the  superfield components
 with  physical on-shell modes  of type IIB supergravity  fields
found earlier in this section.
The  corresponding  monomials in
$\theta_{\L,\R}$ are shown in Tables I,II.
The
 dilaton field
$\phi$  is   the lowest  superfield component,
 while its complex
conjugate $\bar{\phi}$ appears in
 the last component multiplying
$\theta_\R^4\theta_\L^4$.
As another  example,   consider
 the  antisymmetric
2-nd rank complex tensor field modes  $b^\oplus_{ij}$ and
$b_{ij}^\ominus$.
According to Table I, they  correspond to the monomials
$\theta_\L \prp\gamma^{-ij}\theta_\L$ and
$\theta_\R\prm\gamma^{-ij}\theta_\R$ where
we used the  following notation
for the self-dual projectors ($\prp\gamma^{-ij}\equiv
\prp_{ij;kl}\gamma^{-kl}$)
\be
\prp_{ij;kl}=\frac{1}{4}(\delta_{ik}\delta_{jl}-\delta_{il}\delta_{jk}
+\epsilon_{ijkl})\,,\qquad
\prm_{ij;kl}=\frac{1}{4}(\delta_{ik}\delta_{jl}-\delta_{il}\delta_{jk}
-\epsilon_{ijkl})\  . \ee

Let us now determine  the
 equations of motion for the  scalar
superfield $\Phi$.
 For this we  will need the explicit form
of  the second-order Casimir operator for
the   plane
wave superalgebra described in section 2.3

\be\label{cas1} \Q =2P^+P^- + P^IP^I + \f^2 J^{+I}J^{+I}
-\frac{1}{2}\f \bar{Q}^+\Pi\gamma^+ Q^+ \ .  \ee The
representations of the generators of the \pw  superalgebra in
terms of differential operators acting of $\Phi(x,\theta)$ may be
found by using the standard supercoset method (cf.
\rf{kin1}--\rf{kin3})\footnote{In this section
 we use the antihermitean representation for the
generators $P^\mu$. The  corresponding  commutation relations in this
representation can be found from
\rf{pminpi}-\rf{qminqmin} by
 the substitutions $P^\mu \rightarrow -{\rm i}P^\mu$.}

\be
 P^+=\partial^+\,,\qquad \ \ \  P^-=\partial^-\,,
\ \ \ \ \ \ P^I = \cos \f x^+\ \partial^I + \f \sin \f x^+\
x^I\partial^+ \,,\ee \be  J^{+I} =\f^{-1} { \sin \f x^+}\
\partial^I -\cos\f x^+ \ x^I\partial^+\ , \ \ \ \ \ \ \ \ \
J^{IJ} =x^I\partial^J -x^J\partial^I + \
\frac{1}{2}\partial_\theta\gamma^{IJ}\theta  \ , \ee \be  \la{lop}
 Q^+ = -2{\rm
i}\partial^+ \bar{\gamma}^-e^{{\rm i}\f x^+\Pi}\theta \ ,\ \ \ \ \
 \bar{Q}^+ = \frac{1}{2}\bar{\gamma}^-e^{-{\rm
i}\f x^+\Pi}\gamma^+\partial_\theta  \ , \ \ \ \ \ \ \ \
    \{\partial_{\theta^\beta},\theta^\alpha\} =
\frac{1}{2}(\gamma^+\bar{\gamma}^-)^\alpha{}_\beta  \ .  \ee
The projector in the r.h.s. of the definition of the  fermionic
 derivatives $\partial_\theta$ in \rf{lop}
 reflects the fact that $\theta$ satisfies the   light-cone gauge
condition.
Plugging these expressions into \rf{cas1} we find

\be\la{heh} \Q =\Box -2{\rm i}\f
\partial^+\theta\bar{\Pi}\partial_\theta\,,  \ee
where $\Box$ was  defined in \rf{sca}.

In a general curved
background the equations of motion for the superfield $\Phi$
 take the form $({\cal C}-{\cal
C}_0)\Phi=0$,  where the constant term
${\cal C}_0$ should be fixed by an additional
requirement. For example, in the case of the  $AdS$ space, ${\cal
C}_0$ is expressed   in terms of   constant
 curvature of the background.
 In the present case of the  plane wave background
the ${\cal C}_0$ can be fixed by using the so
called ``$sim$'' invariance
 -- the  invariance under the
original plane-wave superalgebra supplemented by the
scale-invariance condition, i.e. by the condition of
 dilatational invariance    in superspace \ci{metp}.\foot{
 In the usual 4 dimensions
 scale transformations (dilatations)
combined with  the Poincare group
 form the maximal subgroup of the
conformal group, or similitude group $SIM(3,1)$.
Dilatation
invariance  ensures masslessness, so the direct generalization to
the supergroup case  should give a criterion of masslessness for
the superfields. }  The generator $\DD$  of dilatations in the  \lc
superspace ($\l=$const) \be \la{ded} \delta x^+ =  0\,, \qquad
\delta x^- = 2\lambda x^-\,, \qquad \delta x^I =  \lambda
x^I\,,\qquad \delta \theta =\lambda \theta\,,\ee has the obvious
form \be \DD = 2x^-\partial^+ +x^I\partial^I +
\theta\partial_\theta\,.\ee The requirement of $sim$ invariance
of the superfield equations of motion   amounts to the condition
$ [{\rm D},{\cal C}]\Phi=0.$ Since, as it is easy to see from
\rf{heh}, $[\DD,\Q]=-2\Q$ it   follows then   that the only
$sim$-invariant equation of motion is simply \be \la{koki} {\cal
C}\Phi=0\,, \ \ \ \ {\rm  i.e.} \ \ \ \ \ \
 (\Box -2{\rm i}\f \partial^+\theta\bar{\Pi}\partial_\theta
)\Phi(x,\theta)=0\,. \ee
The corresponding quadratic term in the
superfield light-cone gauge
action is then

\be S_{l.c.} = \frac{1}{2}\int d^{10}xd^8\theta\,\  \Phi(x,\theta)
(\Box -2{\rm i}\f\partial^+\theta\bar{\Pi}\partial_\theta
)\Phi(x,\theta)\,. \ee
Splitting the  fermionic
coordinate $\theta$
 into $\theta_\R$ and $\theta_\L$ parts  as in \rf{thee}
 one can rewrite \rf{koki} as

\be \label{555}\Bigl[\Box +2{\rm
i}\f\partial^+(\theta_\L\partial_{\theta_\L}
-\theta_\R\partial_{\theta_\R})\Bigr]\Phi(x,\theta_\R,\theta_\L)=0
\ . \ee
This remarkably  simple equation summarizes
all  the field
equations for the
physical fluctuation  modes of type IIB supergravity fields
in the present  R-R \pw background
(i.e. the components of $\Phi$ \rf{supfield})
which were derived earlier in this section.
In particular, the universal expression for
the lowest values of the \lc energy operator can be found by applying
\rf{ccc},\rf{low}
to the case of the equation  \rf{555}:
\be E_0 = \f(4+ \theta_\L\partial_{\theta_\L}
-\theta_\R\partial_{\theta_\R})\,.\ee
This reproduces the values of $\E_0$ in Tables I,II.

\newsection{Concluding Remarks }

In this paper we presented
the quantization of
type IIB string theory in the maximally
supersymmetric R-R \pw background of \ci{bla}
whose \lc gauge action was found in \ci{rrm}.
We explicitly constructed  the quantum \lc Hamiltonian
and the string representation of the corresponding
 supersymmetry algebra.
The  superstring Hamiltonian has the standard
``harmonic-oscillator'' form, i.e.  is quadratic
in creation/annihilation
operators in all 8 transverse directions,
so that its spectrum can be  readily obtained.

We have discussed  in detail the structure of the
 zero-mode sector  of the theory,  giving it the
space-time field-theoretic interpretation  by
 establishing the
precise correspondence  between the lowest-lying
``massless'' string states
and the type IIB supergravity  fluctuation  modes
in the \pw background.

The ``massless''  (supergravity) part of the
  spectrum  has certain similarities
with the supergravity spectrum
  found \ci{kim} in the case of  another maximally supersymmetric
type IIB background -- $AdS_5 \times S^5$ \ci{schwarz}
(this  may not be completely
surprising given that the two backgrounds are related by a
special limit \ci{blah}).
In particular, the     \lc energy
spectrum  of a superstring in the R-R \pw background
is discrete.
As in the $AdS$ case \ci{BF}, the discreteness of the spectrum
 depends
on a particular natural
 choice  of the  boundary conditions. In
 the present case they are  the same as  in the standard harmonic
oscillator problem: the square-integrability of the
wave functions in all  8 transverse spatial directions.

An  interesting feature of the \pw  string spectrum is its
non-trivial dependence on $p^+$. This is  possible due to the
fact  that  the generator $P^+$ commutes with all other
generators of the  symmetry superalgebra. We defined the spectrum
in terms of the \lc energy $H=-P^-$,  which does not depend on
$p^+$ for the  massless (zero-mode) states  but does depend on it
   for the
string oscillator modes. In general, one may define the string
spectrum in curved space
 in terms
of the second-order Casimir operator of the corresponding
superalgebra.  In the present case the eigen-values
of this operator  depend on discrete quantum numbers  as well as
on $p^+$ (through the
dimensionless combination  $\mm= 2 \pi \a' p^+\f$
with the curvature  scale $\f$  and  the string scale $\a'$).

Given the exact solvability of this \pw string theory,
there are many standard  flat-space string
 calculations  that can  be  straightforwardly  repeated in this case.
One can determine the vertex operators for the ``massless''
superstring states and compute the 3-point and 4-point correlation functions,
following the same strategy as
in the \lc Green-Schwarz approach to flat
  superstring theory.\foot{Note that in the present \pw case we
do not have the standard S-matrix set-up:
the string spectrum is discrete
in all 8 transverse
 directions, i.e. the string states  with non-zero
$p^+$ are localized near $x_I=0$
and cannot escape to infinity.}
 It would be interesting to
compare (the $\a' \to 0$ limits of)  the \pw string
results to  the corresponding
correlation functions  in  the  type IIB supergravity on
 $AdS_5\times S^5$.
One can also  find  possible D-brane configurations,
by imposing  open-string boundary conditions in some directions
and repeating the analysis of section 2.\foot{One obvious
candidate  is a D-string  along $x^9$ direction.
For a \lc gauge description of D-branes in flat space see
\ci{GG}.}

\bigskip

Let us  comment on some limits of this  \pw  string theory.
It
depends on  the two mass
 parameters which enter the Hamiltonian \rf{zer}:
the curvature scale $\f$ and the string scale $  (\a' p^+)^{-1} $.
The   limit $\f\to 0$  is the flat-space limit:
the discrete spectrum then becomes   the standard
  type IIB flat-space string  spectrum
 (in the same sense in which  the  harmonic
oscillator spectrum reduces  to the spectrum
of a free particle in the zero-frequency limit).
The $\f\to \infty$ limit is not special:
it corresponds simply to a rescaling  of the \lc energy and
$p^+$ (recall that $\f$ in \rf{bem} can be set to 1 by a rescaling  of
$  x^+ $ and $x^-$).

The limit $\a' p^+ \to 0$   corresponds to the
supergravity  in the \pw background: the string
 Hamiltonian \rf{ttt},\rf{ezer},\rf{eeee}  becomes  infinite
on all states that contain non-zero string oscillators, i.e.
it effectively  reduces to $E_0$ \rf{ezer}  restricted to
the subspace of the zero-mode  states.
The opposite (``zero-tension'') limit
$\a' p^+  \to \infty$   is also regular:
it follows from \rf{zer} that  here we are left with
\be\la{nuu}
H_{\a'p^+ \to \infty}
= \f \bigg[  (a_0^I\bar{a}_0^I +
2\bar{\theta}_0 \bar{\gamma}^-\Pi\theta_0 + 4)
           +   \sum_{\cI=1,2}
 \sum_{n=1}^\infty  (a_n^{\cI I}
\bar{a}_n^{\cI I}+ \eta_n^\cI\bar{\gamma}^-\bar{\eta}_n^\cI ) \bigg]
\ .  \ee
The constraint \rf{cons} remains  the same as it does not involve
$\a'$.
This provides   an interesting example of a
 non-trivial  ``{\it null-string}'' spectrum
which is worth further study. Note, in particular, that
here the energies do not  grow  with
the oscillator level number $n$, i.e. there is no
Regge-type trajectories.\foot{Note that the parameter $\f$
 may  be viewed as a ``regularization''  introduced to define
a non-trivial  tensionless string limit of the flat superstring.}

\bigskip

Let us now compare the \pw string spectrum with the expected form
of the \lc spectrum of the superstring  in \adss background.  In
general, the spectrum of   the \lc Hamiltonian  $\cal H$=  $-{\cal
P}^-$ in \adss  \ci{mtt} should  depend on two
characteristic  mass parameters: the  curvature  scale  $
R^{-1}$  (the inverse $AdS$ radius)\foot{In the context of the
standard AdS/CFT the radius $R$ is related to the `t  Hooft
coupling $\l$ by \ci{hua}
 $R = \lambda^{1/4} \sqrt{\a'}$.}
which is the analog of  $\f$ in \rf{bem}  and  the string  mass scale
 $ \sqrt{\a' } $.
In the context of the standard AdS/CFT correspondence
the coordinates  should be rescaled so that
$R$ is always combined with $\a'$  into the effective
dimensionless tension parameter  $\T = R^2 T = { R^2\ov 2 \pi \a'} =
{ \sqrt \lambda \ov 2 \pi}$.
In contrast to the \pw case,
 here the dependence  of $\cal H$ on
$p^+$   can only be
 the  trivial one,  i.e.  only through the $ { 1 \ov p^+} $
factor (in Poincare coordinates the \adss background has
Lorentz invariance in $(+,-)$ directions).
Let us recall the form  of the \lc string
Hamiltonian
using  the ``conformally-flat'' 10-d  coordinates $(x^a,Z^M)$
 in which
the  \adss  metric is (here $a=0,1,2,3; \ M=1,...,6$)
\be\label{cfc}
ds^2 =R^2 Z^{-2}(dx^a dx^a + dZ^M dZ^M)  \ .
\ee
Splitting the 4-d  coordinates as $x^a= (x^+,x^-,x^\perp)$
and using the appropriate \lc gauge one finds the  following  phase space
 Lagrangian \ci{mtt}
\be
{\cal L} = \PP_\perp\dot{x}_\perp
+ \PP_M \dot{Z}^M
+\frac{{\rm i}}{2}(\theta^i \dot{\theta}_i
+\eta^i\dot{\eta}_i - h.c.) - {\cal  H}  \ , \la{lol} \ee
$$ {\cal H} =
\frac{1}{2 p^+ }\bigg(\PP_\perp^2 +\PP_M\PP_M
+ \T^2 Z^{-4}(\x'_\perp^2+ \Z'^M\Z'^M)
+   Z^{-2}[(\eta^2)^2 + 2 {\rm i}\eta^i \rho^{MN}_{ij} \eta^j  Z_M \PP_N]
  $$  \be \la{zez}
- \  2 \T \Bigl[\ |Z|^{-3}\eta^i \rho_{ij}^M Z^M
(\th'^j - {\rm i}\sqrt{2}|Z|^{-1} \eta^j\x'_\perp)+h.c.\Bigr]\bigg)  \ .
\ee
Compared to \ci{mtt} we have rescaled
the fermions $\theta^i , \eta^i$ ($i=1,2,3,4$)
by $\sqrt { p^+}$ (thus absorbing all spurious  $p^+$-dependence).
$\PP_\perp,\PP_M$ are  the momenta and $\rho^{MN}$
is a  product of Dirac matrices.
 Here   the coordinates and momenta
(including $\cal H$ and $p^+$) are all  dimensionless
(measured in units of $R$), reflecting the rescaling done in \rf{cfc}.
Restoring the canonical  mass dimensions (${\cal H} \to R H$,
$p^+ \to  R p^+$)  the corresponding analog
of the \pw result \rf{zer}  should thus  have the  structure
\be \la{ssi}
H = { 1 \ov p^+ R^2}
 \big[ \E_0  +   \T   \E_{str} (\T)  \big]
=  { 1\ov  p^+} \big[ { 1 \ov  R^2 } \E_0    +   { 1 \ov 2 \pi \a' }
 \E_{str} ( {\textstyle{ R^2 \ov 2\pi \a'}}) \big]
\ ,
\ee
where $\E_0, \E_{str}$ are dimensionless functions
of the parameters and  discrete quantum numbers.

Here the limit $\a'  \to 0$  or $\T\to \infty$  for fixed $  p^+ R^2 $
corresponds to the
type IIB supergravity  \adss background
with only the $\E_0$ part (known
explicitly \ci{kim,mt4})
surviving on the subspace of finite mass states.
The  limit     $R \to \infty$
with fixed $p^+$
should reproduce the   flat space string spectrum
(this  suggests that
 $\E_{str} (\T\to \infty)$ should be  finite).
The limit
 $\T  \to 0$  for fixed $p^+ R^2$
is a  ``null-string'' limit \ci{tee}.
Like the corresponding limit in the \pw case \rf{nuu}
it is expected to be well-defined.

A formal  correspondence  between \rf{ssi}   and
\rf{zer} is established by identifying
$\f$  with ${ 1 \ov p^+ R^2 }$,  so that
$\mm= 2 \pi \a'  p^+ \f  $ in \rf{zer} goes over to  $2\pi \a' \ov R^2$=$\T^{-1}$.
This  rescaling  of $R^2$  by $p^+$  ``explains'' why
\rf{ssi} does not have a non-trivial dependence on $p^+$ while
\rf{zer} does.

The  dependence  of the string-mode part
 $\E_{str}$  of  \rf{ssi}  on $\T$
  should of course  be much more
 complicated than  dependence on  $\a'  p^+ \f $  in \rf{zer}.
To determine it  remains an outstanding problem.

\bigskip

While this  work was nearing completion  there appeared
an interesting paper \ci{malda} which
provides   a gauge-theory interpretation of
this  \pw  string  theory  based on a
 special  limit of the  AdS/CFT correspondence.

\section*{Acknowledgments}
 Thes work of R.R.M. was supported by the
INTAS project 00-00254, by the RFBR Grant 02-02-16944
 and RFBR
Grant  01-02-30024 for Leading Scientific Schools. The
work  of A.A.T. was  supported  in part by the DOE
DE-FG02-91ER-40690, PPARC    PPA/G/S/1998/00613,   INTAS
 991590,  CRDF RPI-2108 grants
and the  Royal Society  Wolfson research merit award.
We are grateful to S. Frolov and  J. Russo  for useful
 discussions.

\setcounter{section}{0} \setcounter{subsection}{0}

\appendix{Notation  and definitions}


We use the following  conventions
for the indices:
\begin{eqnarray*}
 \mun, \nun, \kun =0,1,\ldots 9 && \qquad \hbox{ 10-d  space-time coordinate
indices }
\\
\mu,\nu,\rho = 0,1,\ldots, 9 && \qquad  so(9,1) \  \hbox{ vector
indices (tangent space indices) }
\\
I,J,K,L = 1,\ldots, 8 && \qquad  so(8) \  \hbox{ vector indices
(tangent space indices) }
\\
i,j,k,l = 1,\ldots, 4 && \qquad  so(4) \  \hbox{ vector indices
(tangent space indices) }
\\
\ipr,\jpr,\kpr,\lpr = 5,\ldots, 8 && \qquad  so^\prime(4) \
\hbox{ vector indices (tangent space indices) }
\\
\alpha,\beta,\gamma = 1,\ldots, 16 && \qquad  so(9,1) \  \hbox{
spinor indices in chiral representation}
\\
a,b = 0,1 && \qquad  \hbox{ 2-d world-sheet coordinate indices}
\\
\cI,\cJ = 1,2 && \qquad  \hbox{ labels of the two real MW
spinors}
\end{eqnarray*}
We identify  the transverse target
indices with tangent space indices, i.e.
$x^{\underline{I}} = x^I$,
and avoid  using
the underlined indices in $+$ and $-$ light-cone directions, i.e.
adopt  simplified notation $x^+$, $x^-$. We suppress the flat
space metric tensor $\eta_{\mu\nu}=(-,+,\ldots, +)$ in scalar
products, i.e. $ X^\mu Y^\mu\equiv \eta_{\mu\nu}X^\mu Y^\nu.$
 We decompose
$x^\mu$ into the light-cone and transverse coordinates: $x^\mu=
(x^+,x^-, x^I)$, $x^I=(x^i,x^\ipr)$, where \be x^\pm\equiv
\frac{1}{\sqrt{2}}(x^9\pm x^0)\,.\ee The scalar products of
tangent space vectors are decomposed as \be X^\mu Y^\mu = X^+Y^-
+ X^-Y^+ +X^IY^I\,, \qquad X^IY^I=X^iY^i+X^\ipr Y^\ipr\,. \ee
 The notation $\partial_\pm$, $\partial_I$ is  mostly  used
for target space derivatives\footnote{In sections 1 and 2.1
 $\partial_\pm$ indicate world-sheet derivatives.}
\be
\partial_+ \equiv \frac{\partial}{\partial x^+}
\qquad
\partial_- \equiv \frac{\partial}{\partial x^-} \,,\qquad
\partial_I \equiv \frac{\partial}{\partial x^I}\,.\ee
We also use
\be \partial^+ = \partial_-\,,\qquad
\partial^- = \partial_+\,,\qquad
\partial^I = \partial_I\,.\ee
 The
$SO(9,1)$ Levi-Civita tensor is defined by
$\epsilon^{01\ldots
9}=1$, so that in the light-cone coordinates $\epsilon^{+-1\ldots
8}=1$. The
 derivatives with respect to the  world-sheet coordinates
$(\tau,\sigma)$  are  denoted as
\be \dot{x}^I \equiv \partial_\tau x^I\,, \ \ \ \
 \qquad \x'^I \equiv
\partial_\sigma x^I\,. \ee
We use the chiral representation for the $32\times 32$ Dirac
matrices $\Gamma^\mu$ in terms of the $16\times 16 $
matrices $\gamma^\mu$
\be \Gamma^\mu =\left(\begin{array}{cc} 0  & \gamma^\mu \\
\bar{\gamma}^\mu & 0
\end{array}\right) \,,\ee
\be \gamma^\mu\bar{\gamma}^\nu + \gamma^\nu\bar{\gamma}^\mu
=2\eta^{\mu\nu}\,,\qquad \gamma^\mu =
(\gamma^\mu)^{\alpha\beta}\,, \qquad  \bar{\gamma}^\mu
=\gamma^\mu_{\alpha\beta}\,, \ee \be
\label{gammaspl}\gamma^\mu=(1,\gamma^I,\gamma^9)\,,\qquad
\bar{\gamma}^\mu=(-1,\gamma^I,\gamma^9)\,,\qquad
\alpha,\beta=1,\ldots 16\,.\ee We adopt the Majorana
representation for $\Gamma$-matrices, $ C= \Gamma^0$, which
implies that all $\gamma^\mu$ matrices are real and symmetric,
$\gamma^\mu_{\alpha\beta} = \gamma^\mu_{\beta\alpha}$,
$(\gamma^\mu_{\alpha\beta })^* = \gamma^\mu_{\alpha\beta}$. As in
\ci{rrm} $\gamma^{\mu_1\ldots \mu_k}$ are the antisymmetrized
products of $k$ gamma matrices, e.g.,
$(\gamma^{\mu\nu})^\alpha{}_\beta \equiv
\frac{1}{2}(\gamma^\mu\bar{\gamma}^\nu)^\alpha{}_\beta -(\mu
\leftrightarrow \nu)$, $ (\gamma^{\mu\nu\rho})^{\alpha\beta}
\equiv
\frac{1}{6}(\gamma^\mu\bar{\gamma}^\nu\gamma^\rho)^{\alpha\beta}
\pm 5 \hbox{ terms}$. { Note that
$(\gamma^{\mu\nu\rho})^{\alpha\beta}$ are antisymmetric in
$\alpha$, $\beta$.}
 We assume the normalization
\be\label{g11} \Gamma_{11} \equiv  \Gamma^0\ldots \Gamma^9
=\left(\begin{array}{cc}
1 & 0\\
0 & -1
\end{array}\right)\ , \ \ \ \ \ \ \ \ \
 \gamma^0\bar{\gamma}^1
\ldots \gamma^8\bar{\gamma}^9=I \ .
\ee
We use the following definitions
\be \Pi^\alpha{}_\beta \equiv
(\gamma^1\bar{\gamma}^2\gamma^3\bar{\gamma}^4)^\alpha{}_\beta\,,\qquad
 (\Pi^\prime)^\alpha{}_\beta \equiv
(\gamma^5\bar{\gamma}^6\gamma^7\bar{\gamma}^8)^\alpha{}_\beta\,.\ee
\be \bar{\Pi}_\alpha{}^\beta \equiv
(\bar{\gamma}^1\gamma^2\bar{\gamma}^3\gamma^4)_\alpha{}^\beta\,,\qquad
 (\bar{\Pi}^\prime)_\alpha{}^\beta \equiv
(\bar{\gamma}^5\gamma^6\bar{\gamma}^7\gamma^8)_\alpha{}^\beta\,.\ee
{Note that
$\Pi^\alpha{}_\beta=\bar{\Pi}_\beta{}^\alpha$}.
Because of the
relation $\gamma^0\bar{\gamma}^9 =\gamma^{+-}$ the normalization
condition \rf{g11} takes the form $ \gamma^{+-}\Pi\Pi^\prime =
1$. Note also the following useful relations (see also \ci{rrm})
\be (\gamma^{+-})^2 = \Pi^2 =(\Pi^\prime)^2 =1\,,\ee \be
\gamma^{+-}\gamma^{\pm} =\pm \gamma^\pm\,,\qquad \bar{\gamma}^\pm
\gamma^{+-} = \mp \bar{\gamma}^\pm\,,\qquad
\gamma^+\bar{\gamma}^+ = \gamma^-\bar{\gamma}^- =0\,,\ee \be
\bar{\gamma}^+(\Pi+\Pi^\prime)=(\Pi+\Pi^\prime) \gamma^- =0\,,
\qquad \bar{\gamma}^-(\Pi - \Pi^\prime)= (\Pi - \Pi^\prime)
\gamma^+ =0\,. \ee \be  \gamma^\pm \bar{\Pi} = \Pi
\gamma^\pm\,,\quad \gamma^i \bar{\Pi} = -\Pi\gamma^i,\quad
\bar{\gamma}^i\Pi = - \bar{\Pi}\bar{\gamma}^i,\quad
\gamma^i\bar{\Pi}' = \Pi'\gamma^i\,, \quad \bar{\gamma}^i\Pi' =
\bar{\Pi}'\bar{\gamma}^i\,.\ee
 The 32-component positive chirality spinor
$\theta$ and  the negative chirality spinor $Q$  are decomposed
in terms of the 16-component spinors as
\be \theta = \left( \begin{array}{c} \theta^\alpha \\
0\end{array}\right)\,, \qquad\quad
Q = \left( \begin{array}{c} 0 \\
Q_\alpha\end{array}\right)\,.\ee
The   complex
Weyl spinor $\theta$ is related to the  two real
Majorana-Weyl spinors $\theta^1$ and $\theta^2$  by
\be\label{comrea} \theta = \frac{1}{\sqrt{2}}(\theta^1 +{\rm
i}\theta^2)\,,\qquad \bar{\theta} = \frac{1}{\sqrt{2}}(\theta^1 -
{\rm i}\theta^2)\,.\ee
The short-hand notation like
$\bar{\theta}\bar{\gamma}^\mu\theta$ and $\bar{\gamma}^\mu\theta$
 stand for
$\bar{\theta}{}^\alpha\gamma_{\alpha\beta}^\mu\theta^\beta$ and
$\gamma_{\alpha\beta}^\mu \theta^\beta$ respectively.


\end{document}